\documentclass[useAMS,usenatbib,usegraphicx]{mn2e}
\usepackage{epsfig}
\usepackage{amsmath} 
\usepackage{rotating}           
\usepackage{color}     
\usepackage{graphicx}
\usepackage{times}
\usepackage{upgreek} 
\usepackage{multicol}

\def\Mo{M$_\odot$}

\def\ccm {$\hbox{{\rm cm}}^{-3}$}    
\def\scm  {$\hbox{{\rm cm}}^{-2}$}
\def \AL {$\alpha $}     
\def \HI {H{\sc \,i}}
\def \HII {H{\sc \,ii}}

\def\lapp{\ifmmode\stackrel{<}{_{\sim}}\else$\stackrel{<}{_{\sim}}$\fi}
\def\gapp{\ifmmode\stackrel{>}{_{\sim}}\else$\stackrel{>}{_{\sim}}$\fi}

\title[Radio turnover and ionsing photons]{The relationship between the turnover frequency and photo-ionisation in radio sources}

\author[S. J. Curran]{S. J. Curran\thanks{Stephen.Curran@vuw.ac.nz}\\
School of Chemical and Physical Sciences, Victoria University of Wellington, PO Box 600, Wellington 6140, New Zealand}

\begin{document}

 \date{Accepted ---. Received ---; in original form ---}

\pagerange{\pageref{firstpage}--\pageref{lastpage}} \pubyear{2024}

\maketitle
\label{firstpage}
\begin{abstract}
  We investigate the connection between the turnover frequency in the
  radio spectrum, $\nu_{\rm TO}$, and the rate of ionising
  ultra-violet photons, $Q_{\text{\HI}}$, in extragalactic sources.
  From a large, optically selected, sample we find $\nu_{\rm TO}$ to
  be correlated with $Q_{\text{\HI}}$ in sources which exhibit a
  turnover. The significance of the correlation decreases when
  we include power-law radio sources as limits, by assuming that the
  turnover frequency occurs below the lowest value observed. However,
  the power-law fit sources are less well sampled across the
  band and so these may just be contributing noise to the data. 
  Given that the observed $\nu_{\rm TO}$--$Q_{\text{\HI}}$
  correlation is purely empirical, we use the ionising photon rate to
  obtain the electron density in a free-free absorption model. For
  each of the constant, exponential, constant plus exponential (Milky
  Way) and spherical models of the gas distribution, there is also an
  increase in the turnover frequency with ionising photon rate.
  Furthermore, for a given gas mass, we find that the turnover
  frequency is anti-correlated with the scale-factor of the gas
  density.  While other mechanisms, such as ageing electrons or
  synchrotron self-absorption, may be required to reproduce the
  spectral indices, for an exponential scale-factor similar to the
  linear size, this simple free-free absorption model reproduces the
  turnover-- size correlation seen in radio sources.
  \end{abstract}  

\section{Introduction} 
\label{intro}

High-frequency peaked/gigahertz peaked-spectrum (HFP/GPS) and compact
steep-spectrum (CSS) sources form a class of radio source which
exhibit a peak in their radio spectrum, while the emission appears to
be confined to a small region.  Specifically HFPs and GPSs have
projected linear sizes of $\ell\lapp1$~kpc and peak in their
luminosity at a turnover frequency of $\nu_{\rm TO} \sim0.4 - 5$~GHz,
whereas CSSs have $\ell\gapp10$~kpc and $\nu_{\rm TO}\lapp0.5$~GHz.
This reflects the well established anti-correlation between $\nu_{\rm TO}$
and $\ell$  \citep{ffs+90,ob97,ode98}.

The main competing hypotheses for the nature of these are
\begin{enumerate}
\item that the more compact sources are young and will evolve into the larger CSSs \citep{ffd+95},
\item that all of the sources are due to confinement of the radio jets by
a dense interstellar medium (the ``frustrated'' model, \citealt{obs91}),
\item or a combination of both (e.g. \citealt{bdo97,beg99,ceg+17}).
\end{enumerate}
The main issue with the youth argument is the apparent under-abundance
of the larger active galactic nucleus (AGN)  population into which they canonically evolve
\citep{rtpw96,ob97,ab12}. However, there is mounting evidence that, in
addition to frustrated sources and those which will evolve, 
many of the ``young'' sources are transient with their radio activity
being cyclic on a time-scale of $\sim1000$~yr
\citep{ab12,cge+15,ceg+17,os21}.

The large rotation measure in some compact radio sources suggests the
presence of large magnetic fields (e.g. \citealt{ktia87}), indicating
that the radio emission is synchrotron in nature
(e.g. \citealt{bdo97}). However, the majority show only weak
polarisation \citep{sod+98}, which may therefore favour free-free
absorption as the dominant mechanism
(e.g. \citealt{mpr+14,tmc+15,kcr19}), although \citeauthor{bdo97}
suggest that a large number of magnetic field reversals could keep the
net polarisation low.

The dominance of the electron density (free-free absorption) is supported by
anti-correlation between the neutral gas abundance, as
traced through absorption of the 21-centimetre transition of hydrogen (\HI), and
turnover frequency in 196 $z\geq0.1$ radio sources \citep{chj+19}:
The detection rate of  \HI\ 21-cm absorption is also strongly
anti-correlated with the photo-ionisation rate, $Q_{\text{\HI}}$, \citep{cww+08,cw12}. where
$Q_{\text{\HI}}$ drives the electron density. 

Thus, we were interested to test whether any of the nine $z=5.6-6.6$
quasars, in which only one had evidence of a possible turnover
\citep{gsi+23}, have relatively low ionising photon rates. 
However, sufficient rest-frame ultra-violet photometry
was available for only one source, which did in fact have an ionising
rate well below the cut-off where neutral gas
has ever been detected ($Q_{\text{\HI}}\approx3\times10^{56}$~s$^{-1}$).
This is therefore consistent with the argument that no turnover is evident due to a 
low ionisation rate and, thus, a low electron density.

However, given that this is a single source, to ensure a sample with
more comprehensive ultra-violet data, we also  searched the photometry
of a larger, optically selected sample.
From this we find a positive correlation between the turnover
frequency and ionising rate, which is supported by a theoretical
model. The model also yields an anti-correlation between the turnover
frequency and the scale-length of the gas density, which could
naturally explain the observed $\nu_{\rm TO}$--$\ell$ anti-correlation.

\section{The observed \boldmath{$\nu_{\rm TO}-Q_{\text{\HI}}$} correlation}
\subsection{Photometry and fitting}
\label{pandf}

We follow the same procedure as described in \citet{cwsb12}, where the data were
scraped from the  {\em NASA/IPAC Extragalactic Database} (NED)\footnote{See Appendix A.},
the {\em Wide-Field
 Infrared Survey Explorer} (WISE, \citealt{wem+10}) {\em Two
  Micron All Sky Survey} (2MASS, \citealt{scs+06}) and the
{\em Galaxy Evolution Explorer} (GALEX, data release
GR6/7)\footnote{http://galex.stsci.edu/GR6/\#mission} databases.
For the \citet{gsi+23} sources the radio data from NED were supplemented with 54~MHz to 3~GHz
photometry from \citet{ceg+17,scr+22,gds+22}.
After shifting the data back into the source's rest-frame,
each flux density measurement, $S_{\nu}$, was then converted to a specific luminosity, via
$L_{\nu}=4\pi \, D_{\rm L}^2 \,S_{\nu}/(z+1)$, where $D_{\rm L}$ is the luminosity
distance to the source.\footnote{We use $H_{0}=67.4$~km~s$^{-1}$~Mpc$^{-1}$ and
$\Omega_{\rm m}=0.3125$ \citep{paa+20} throughout the paper.}

\begin{figure}
  \centering\includegraphics[scale=0.57]{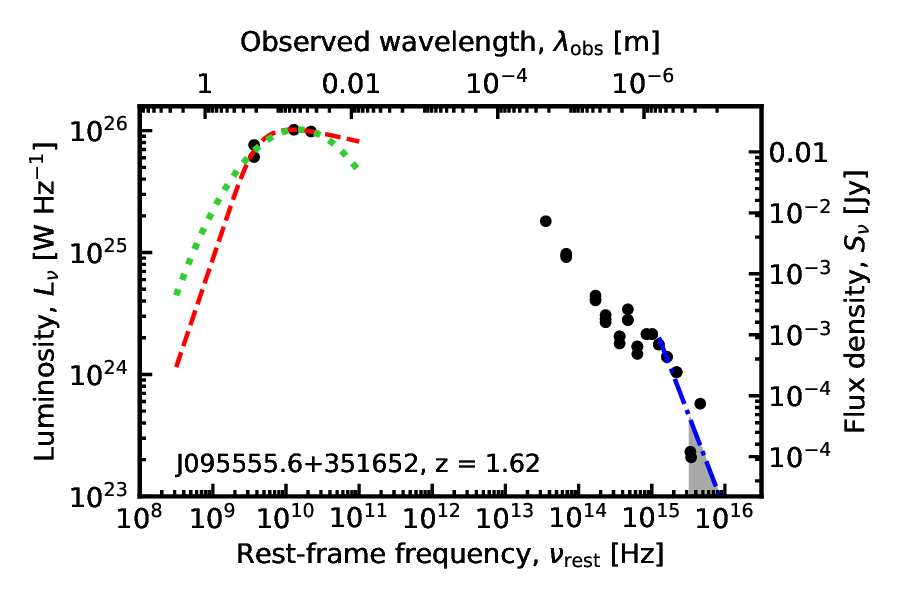}
 \vspace*{-0.6cm}
 \caption{Example of the photometry fitting (from the SDSS sample, Sect.~\ref{paf}).
   The radio fits are shown as the dotted (polynomial) and dashed (GPS-fit) curves
   and the UV as the power-law fit. The shaded region shows $\nu \geq
  3.29\times10^{15}$~Hz, over which the ionising photon rate is
  calculated.  For the GPS-fit shown $\alpha_{\rm thick} = 1.77$ and $\alpha_{\rm thin} = -0.14$. }
\label{Ex}
\end{figure}

After the removal of duplicate measurements at exactly the same frequency, 
where there were at least three radio measurements, we applied both log-space first and second order
polynomial fits to the data, selecting that which gave the lowest sum residual,
$\sum (L_{\text{data}} - L_{\text{fit}})^2$, Fig.~\ref{Ex}. If there were less than
three measurements the source was not used.
Note that the 2nd order fit sources which gave a turnover frequency below the minimum observed
were re-fitted with a power law.
We used the values from the polynomial as the initial guesses to a
GPS-fit function (equation 1 of \citealt{ssd+98}).  However, due to
the limited number of well separated radio photometry points, the bunching of
the data at similar frequencies gave  much more disparate fits.
The polynomial fitting appeared to be more robust, mostly giving reasonable agreement with the
GPS-fits, and better agreement with those of \citet{ob97}.\footnote{See Appendix B.}

The ionising ($\lambda\leq912$~\AA) photon rate was determined from the 
rest-frame UV luminosities, via \citep{ost89}
\begin{equation}
Q_\text{\HI}\equiv\int^{\infty}_{\nu_{\rm ion}}\frac{L_{\nu}}{h\nu}\,d{\nu},
\end{equation}
where $\nu$ is the frequency (with $\nu_{\rm ion} = 3.29\times10^{15}$~Hz for \HI) and $h$ the Planck constant.
Again, requiring at least three photometry measurements,
fitting the rest-frame UV data with a power-law fit, $L_{\nu} \propto \nu^{\alpha}$, gives
\begin{equation}
\log_{10}L_{\nu} = \alpha\log_{10}\nu+ {\cal C},
\end{equation}
where $C$ is the log-space intercept and $\alpha$ the gradient (the UV spectral index,
see Fig.~\ref{Ex}).
Integrating this over $\nu_{\rm ion}$ to $\infty$ gives the ionising photon rate as
\begin{equation}
  Q_\text{\HI} = \frac{-10^{\cal C}}{\alpha h}\nu_{\rm ion}^{\alpha},
  \label{Qcalc}
\end{equation}
which was then calculated for each source with sufficient rest-frame UV photometry.

\subsection{Samples}
\subsubsection{The \citeauthor{gsi+23} sample}
\label{sect:J0002+2550}

As stated above, only one source, J0002+2550
(WISEA\,J000239.39+255035.1), had sufficient rest-frame UV data to
determine the ionising photon rate.\footnote{Of the remaining eight, only four had any rest-frame
IR–UV photometry and are shown in Appendix C.}
\begin{figure}
\centering\includegraphics[scale=0.57]{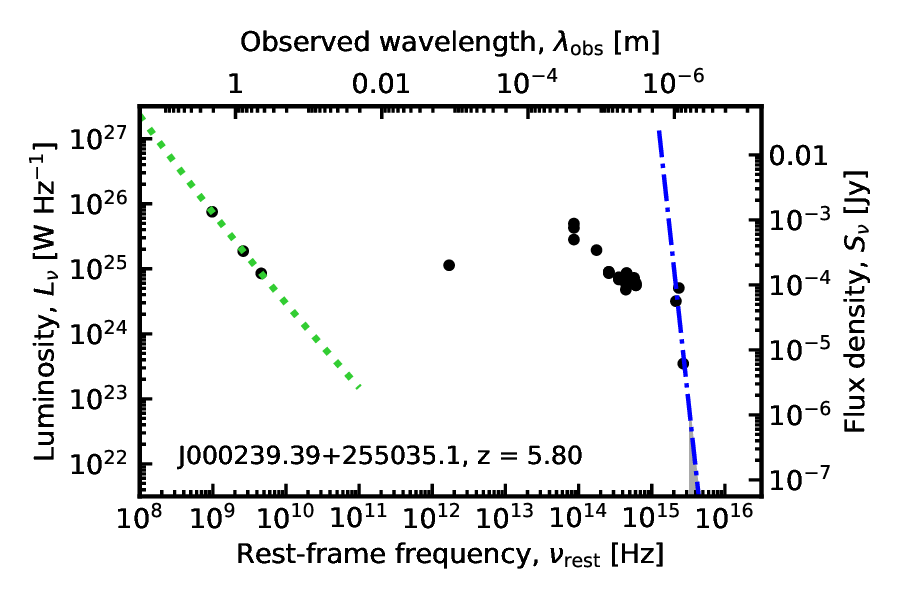}
\vspace*{-0.3cm}
\caption{The rest-frame  photometry of the one source of
  \citeauthor{gsi+23} which had sufficient UV data to yield
  $Q_{\text{\HI}}$. These are from the Pan-STARRS1 $i,z,y$ bands \citep{bvm+14} and
the radio data from \citet{gsi+23}.}
\label{J0002}
\end{figure}
This is shown in Fig.~\ref{J0002}, from which we obtain an ionising
rate of $Q_{\text{\HI}} = 9.3\times10^{54}$~s$^{-1}$.  This is well
below the $\approx3\times10^{56}$~s$^{-1}$ limit at which all of the
gas in the host galaxy is believed to be ionised \citep{cw12} and well
within the regime where \HI\ 21-cm absorption has been detected
\citep{cww+08,chj+19}.

Thus, for the one source for which we could obtain the photometry, the
absence of a turnover in the radio SED is consistent with a relatively
low ionisation fraction. We do note, however, that this has an
unusually steep spectral index in the UV band ($\alpha_{\rm UV}= -10.37$)
and whether this applies to the other eight sources in the
sample we cannot comment. At such high redshifts we would expect a
steepening of the spectral index due to intervening hydrogen,
specifically the Lyman-\AL\ forest (e.g. \citealt{jbr02,bc10}) and
Lyman Limits systems (e.g. \citealt{mvw10}).  In Fig.~\ref{J0002},
$\lambda = 1216$~\AA\ occurs immediately after the second highest
frequency point and so the above ionising photon rate should be
treated as a lower limit (see Sect.~\ref{TOJ0002+2550}).

In order to investigate the relation between the
ionising photon rate and turnover frequency, we turned to a large
sample, which is optically selected in order to increase the
likelihood of yielding the necessary rest-frame UV photometry.

\subsubsection{The SDSS sample} 
\label{paf}

We use the sample of \citet{cmp21}, which comprises the first 100\,000
QSOs of the SDSS Data Release 12 (DR12, \citealt{aaa+15}) with
accurate redshifts. From these we shortlisted the sources which had at
least one flux measurement below 10~GHz, giving 3429 radio sources.
As stated above, we required at least three flux measurements to fit
the radio data, which gave 1005 sources of which 499 could be fit. Of these, 
the ionising photon rate could be obtained for 416 source, of which 
257 sources which were fit by a power law with 159 exhibiting  a turnover.
Since we are primarily interested in the turnover frequency, we used the polynomial results
in the following analysis.

\begin{figure}
  \centering\includegraphics[scale=0.57]{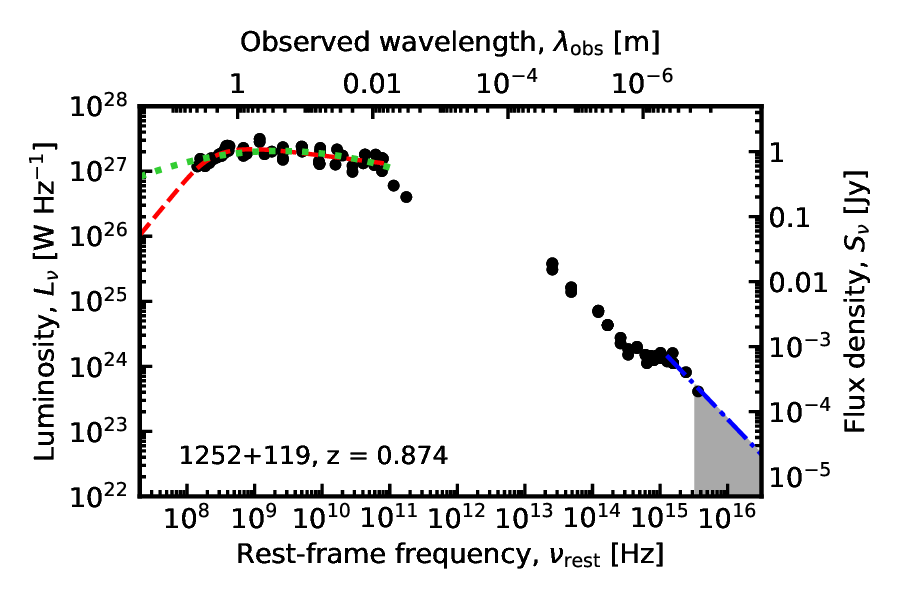}
\vspace*{-0.6cm}
\caption{The rest-frame photometry of the one SDSS source which was common to the low
  frequency catalogues (from the peaked-spectrum sources of \citealt{ceg+17}).}
\label{1252}
\end{figure}

In order to supplement the radio data, we added photometry compiled in surveys
with the {\em GaLactic and Extragalactic All-sky Murchison Widefield Array} (GLEAM),
specifically:
\begin{enumerate}
\item The 1483 sources with  $72 \leq \nu_{\rm TO} \leq 1400$~Mz \citep{ceg+17}.
\item  The 373 sources of \citet{scr+22}, of which 36 exhibit $\nu_{\rm TO} \sim 150$~Mz.
  \item The 24 high redshift ($4.9 \leq z \leq 6.6$) radio bright sources of \citet{gds+22}.
\end{enumerate}
Of the first two samples, spectroscopic redshifts were available for
214 and 49 sources, respectively. Comparing these with the SDSS sample
gave just a single match, \protect[HB89]\,1252+119
(GLEAM\,J125438+114103) at $z=0.872$.\footnote{Note
that there are two other coordinate matches, \protect[HB89]\,1307+121
(GLEAM\, J130934+115425) and \protect[HB89]\,1442+101
(GLEAM\,J144516+095835), although the redshifts of the SDSS and GLEAM
sources are very different -- $z = 2.6006$, cf. 0.34286 and
$z=3.5203$, cf. 0.25578, respectively.} The lowest observed rest-frame
frequency is 141~MHz, compared to 676~MHz for the NED only data.
For the low frequency plus NED photometry, the 
polynomial fit gives $\nu_{\text{TO}}=10^{9.36\pm0.03}$~Hz = $2.29_{-0.07}^{+0.23}$~GHz and
the GPS fit 251~MHz (Fig.~\ref{1252}), whereas the NED only data gives
$\nu_{\text{TO}}=10^{8.40\pm1.70}$~Hz = $252_{-247}^{+12400}$~MHz and the GPS fit 603~MHz\footnote{Which, being
lower than the minimum observed frequency, would be re-fitted with a power law.}.
This is actually closer to the $\nu_{\text{TO}}=608$~MHz of \citet{ceg+17}\footnote{325~MHz in the observed
frame (J. Callingham, private communication).} which uses the
low frequency data.

\subsubsection{The low frequency samples} 

We also performed full photometry searches on the low frequency
samples.  Of the sources with
at least three radio photometry
measurements, 260 have spectroscopic redshifts, of which 24 were best
fit by a power law and 201 with a turnover.\footnote{Although the
remaining 35 appeared to be amply sampled, some gave inverted spectra and
the bunching of data at similar frequencies in others prevented a fit.} Just 55 of these had sufficient
rest-frame UV photometry to obtain the ionising photon rate and adding
these to the SDSS sources with a measured $Q_{\text{\HI}}$, gave 263
first order fit sources and 208 second order fit.

\subsection{Selection effects}

\subsubsection{Radio flux limits}
\label{rfl}

\begin{figure}
\includegraphics[scale=0.52]{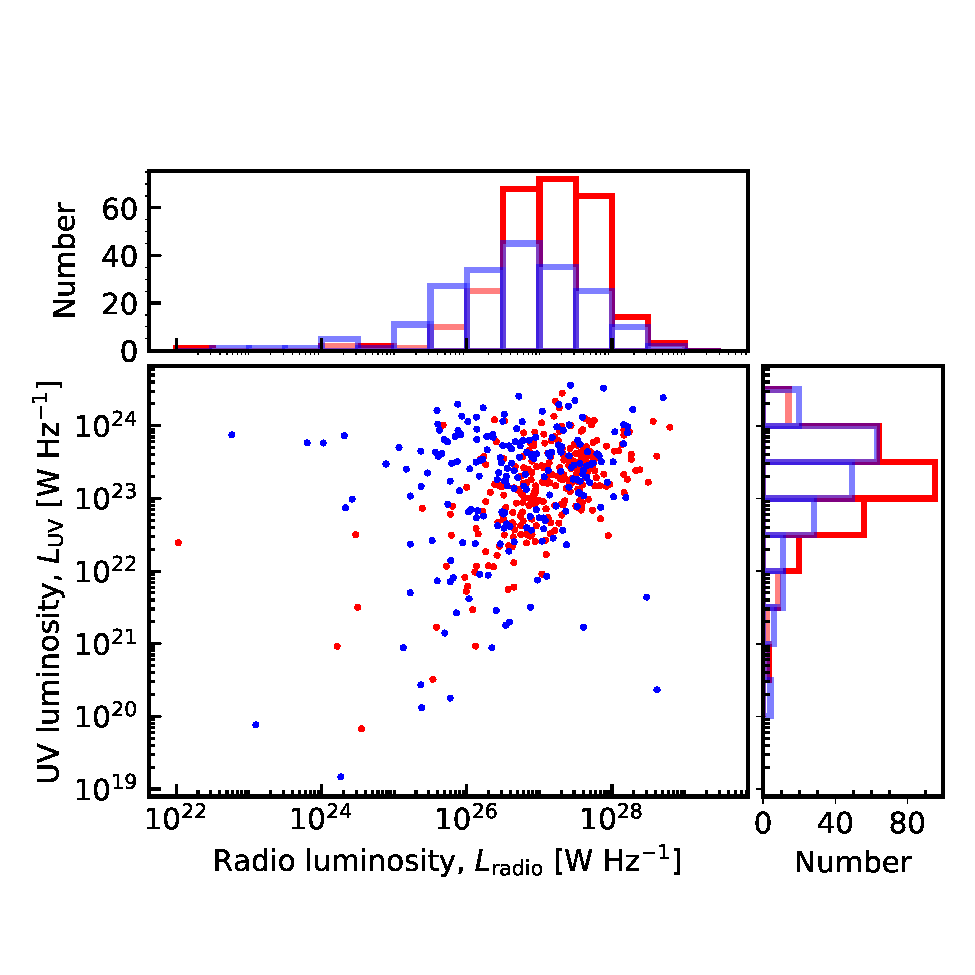}
   \vspace*{-1.2cm}
   \caption{The UV ($\lambda = 912$~\AA) and radio ($\lambda = 21$~cm)
     luminosities of the SDSS plus low frequency samples. For the
     1st order fit sources (red markers/histogram) the mean luminosities are
     $\left<\log_{10}L_{\text{radio}}\right> = 27.08\pm0.75$ and
     $\left<\log_{10}L_{\text{UV}}\right> = 23.15\pm0.65$ and for the 2nd order (blue
    markers/histogram)
     fit sources, $\left<\log_{10}L_{\text{radio}}\right> = 26.61\pm0.98$ and
     $\left<\log_{10}L_{\text{UV}}\right> = 23.14\pm0.93$.  A $t$-test gives a probability of
     $p(t) = 9.61\times10^{-9}$  that the mean radio luminosities are the
     same and $p(t) = 0.911$ that the mean UV luminosities are the same.}
\label{radio_UV}
\end{figure}

In Fig.~\ref{radio_UV}  we show the radio and UV luminosity distributions of the
combined SDSS plus low frequency sample.
The UV and radio luminosities are strongly correlated, with a
Kendall-tau test giving a probability of $p(\tau) =
9.20\times10^{-13}$ of this arising by chance, which is significant at
$Z(\tau) =7.14\sigma$, assuming Gaussian statistics.  Also, it
is apparent that the 1st order fit sources are generally more luminous
at $\lambda = 21$~cm than those with a  2nd order fit.

\begin{figure}
\includegraphics[scale=0.57]{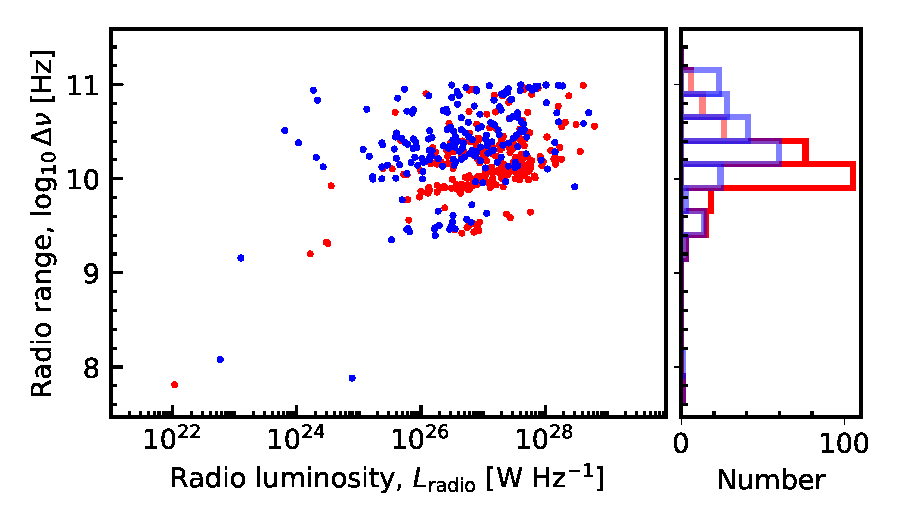}
   \vspace*{-0.5cm}
   \caption{The frequency spans of the SDSS plus low frequency
     samples, defined as $\log_{10}\Delta\nu = \log_{10}(\nu_{\text{max}} - \nu_{\text{min}})$
     up to our 100~GHz cut.
     For the 1st order fit (red markers/histogram) sources
     $\left<\log_{10}\Delta\nu\right> = 10.15\pm0.35$ and for the 2nd
     order fit (blue markers/histogram) sources $\left<\log_{10}\Delta\nu\right> =
     10.35\pm0.47$.  A $t$-test gives a probability of $p(t) =
     9.55\times10^{-8}$ ($5.34$) that the mean sampled frequency spans are the
     same.}
\label{radio_span}
\end{figure}

This could be due to the flagging of 2nd order fits as 1st order
if the fitted turnover frequency is below the minimum observed, 
which will have the effect of boosting the luminosity if
the putative turnover occurs at
$\nu_{\text{TO}} < 1.4$~GHz. 
This may be evident in  Fig.~\ref{radio_span}, where 
the 2nd order fit sources are, on average, more fully sampled,
thus making it possible that some of the 1st order fit sources may
exhibit a turnover if sampled to a similar degree.
The median radio bandwidths of the low frequency observations is 
$\Delta\nu\sim750$~MHz \citep{ceg+17,scr+22} and
so we remove our most extreme outliers (at $\Delta\nu\lapp1$~GHz) from the sample in the following
analysis.

\subsubsection{UV photometry correction}
\label{uvsi}

\begin{figure}
\hspace*{-0.3cm}
\centering\includegraphics[scale=0.50]{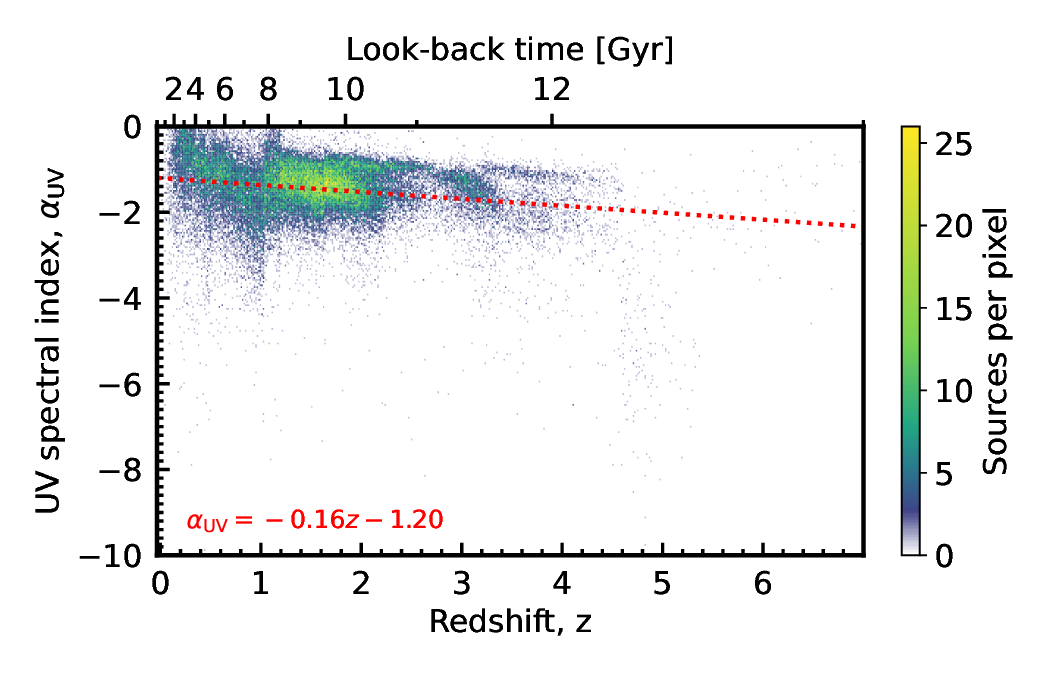}
\vspace*{-0.7cm}
\caption{The distribution of the UV spectral index with redshift for
  the SDSS sources for which the ionising photon rate could be
  obtained. The dotted line shows the best linear fit.}
\label{UV_alpha}
\end{figure}

\begin{figure}
\hspace*{-0.3cm}
\centering\includegraphics[scale=0.5]{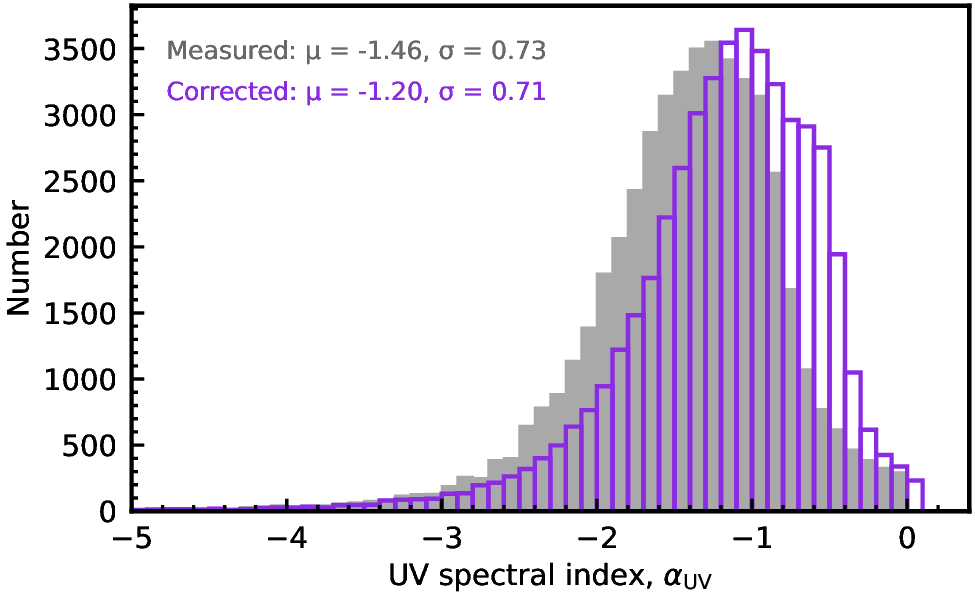} 
\vspace*{-0.2cm}
\caption{The distribution of the UV spectral index before correction (unfilled histogram)
and after correction (filled histogram).} 
\label{UV_alpha_hist}
\end{figure}

As discussed in Sec.~\ref{sect:J0002+2550}, the higher the redshift the more we
expect the rest-frame UV continuum to be absorbed by intervening hydrogen.
\begin{figure}
  \includegraphics[scale=0.53]{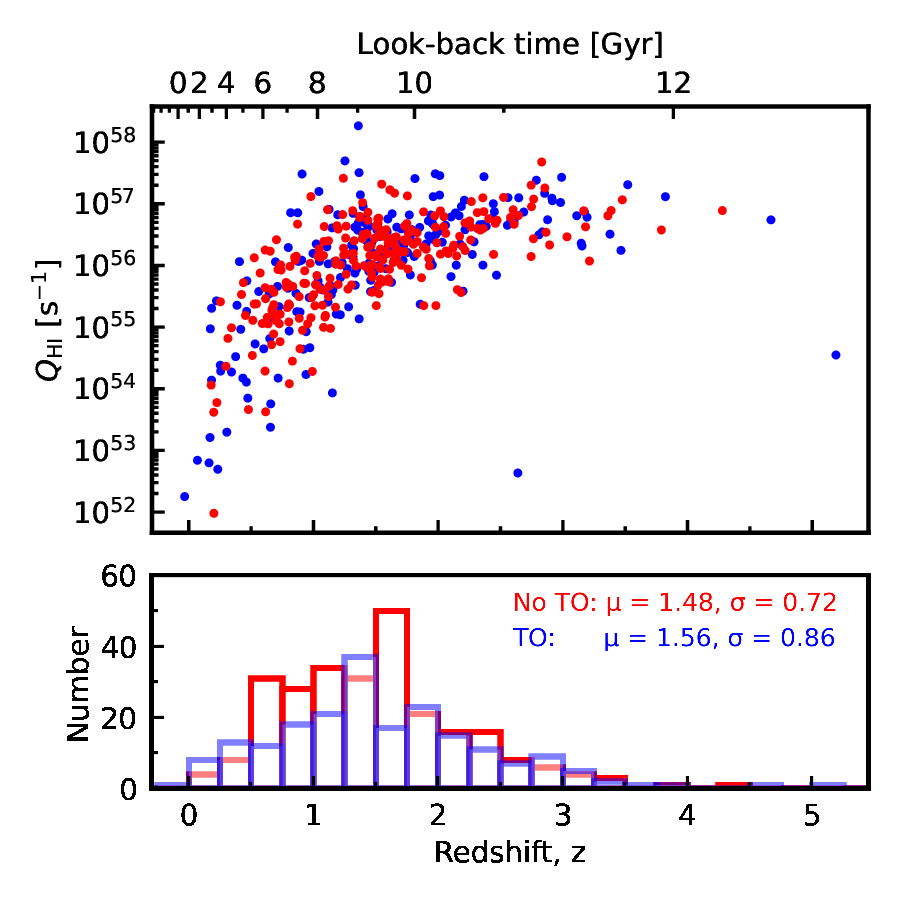}
  \vspace*{-0.4cm}
  \caption{The ionising photon rate versus redshift for the SDSS radio
    plus low frequency source samples. The red markers/histogram are
    the power-law ({\sf No TO}) radio sources and the blue
    markers/histogram those exhibiting a turnover ({\sf TO}). The
    mean values and associated standard deviations are given in the
    lower panel. A $t$-test gives a probability of
     $p(t) = 0.271$ that the mean redshifts are the same.}
\label{SDSS_z-Q}
\end{figure}
In the top panel of Fig.~\ref{SDSS_z-Q}, we can see a clear Malmquist bias, where the volume
probed at low redshift is insufficient to detect the rare high luminosity sources,
while at high redshift only the high luminosity sources are detected.
We therefore use the spectral index to quantify a correction for the intervening
hydrogen. In Fig.~\ref{UV_alpha} we show all of the SDSS sources for which we could
obtain an ionising photon rate,
from which we can see a steepening in the spectral index with
redshift. A linear fit gives $\alpha_{\rm UV} = -0.16z-1.20$, so, on
the basis that the intrinsic spectral index is $\alpha_{\rm UV} =
-1.20$ at $z\sim0$, we correct each measured value of $\alpha_{\rm
  UV}$ by subtracting $-0.16z$
(Fig.~\ref{UV_alpha_hist}).\footnote{$\alpha_{\rm UV} = -1.20$ at
$z\sim0$ is similar to values in the literature, e.g.  $\alpha_{\rm
  UV} =-1.4$ at $z<1.44$ \citep{ssd12} and $\alpha_{\rm UV} = -1.7$ at
$\left<z\right> =1$ \citep{tzkd02}.}
Although not valid for an individual source, using this to correct the
UV spectral index to obtain the intrinsic value provides a  statistical correction for the
redshift, and thus a better estimate of the ionising photon rate before the
absorption of the UV continuum by intervening gas.\footnote{There is also the issue
of dust suppressing the shorter wavelengths in the rest-frame, leading to underestimates in
$Q_{\text{\HI}}$. However, intrinsic dust will also provide shielding for the gas countering the degree of 
ionisation somewhat.}

\subsection{Ionising photon rate and turnover frequency}
\subsubsection{Correlations}
\label{sec:res}

\begin{figure}
  \hspace*{3mm} \includegraphics[scale=0.48]{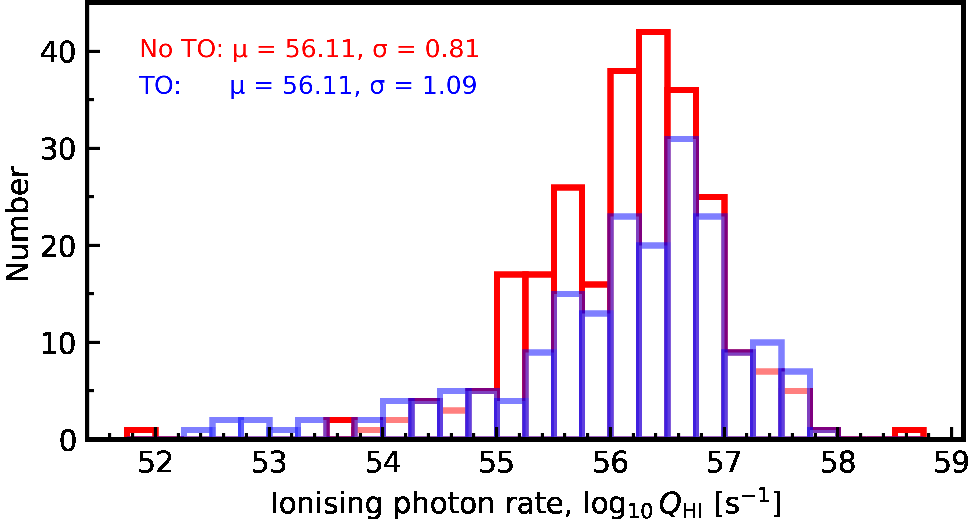}
   \vspace{-2mm}
   \caption{The distribution of the corrected ionising photon rates
     for the radio sources fit by a power-law ({\sf No TO}) and
     exhibiting a turnover ({\sf TO}).}
\label{Q_histo}
\end{figure}

From the corrected spectral indices we recalculate the ionising photon rates, according to
Equ.~\ref{Qcalc}. Showing these in  Fig.~\ref{Q_histo},
we see little difference between the 1st and 2nd order fit sources.
A $t$-test gives a probability of $p(t) = 0.952$
of the mean $Q_{\text{\HI}}$ values being the same and a Kolmogorov-Smirnov test gives $p(KS) =
0.032$ of the populations being drawn from
the same sample.\footnote{For the uncorrected ionising photon rates
these are $p(t) = 0.992$ and $p(KS)=0.045$.}

\begin{figure}
  \includegraphics[scale=0.55]{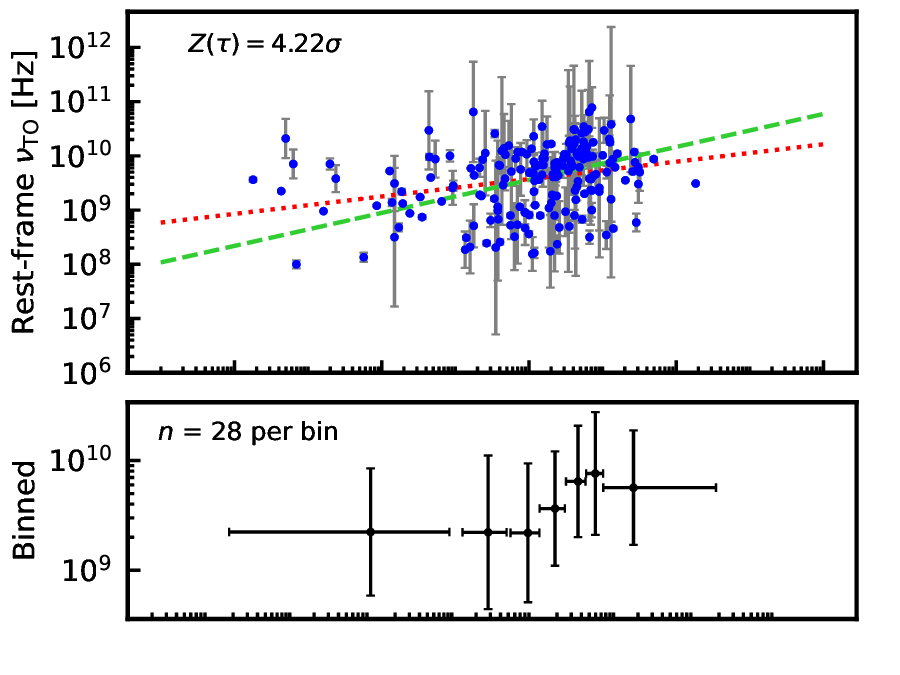} 
  \vspace*{-0.6cm}
  
  \includegraphics[scale=0.55]{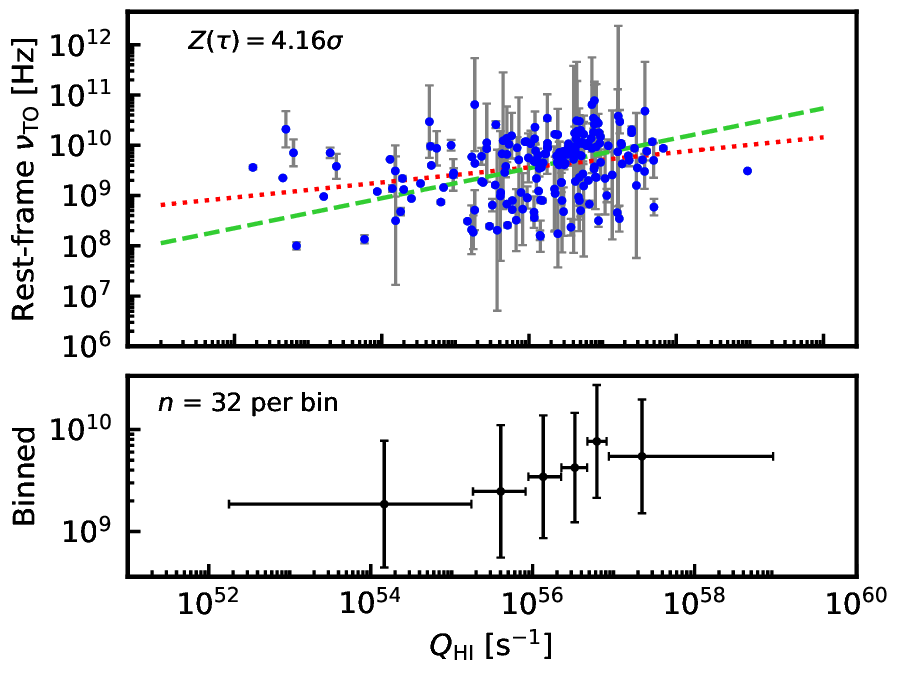}
   \vspace*{-0.6cm}
  \caption{The rest-frame turnover frequency versus the measured (top)
    and corrected (bottom) ionising photon rate.
    In the top panel of each subplot the error bars show the uncertainty in the turnover
    frequency from the covariance of the 2nd order fit. The dotted (red) line shows the least-squares
    linear fit and the dashed (green) line the fit weighted by the $1\sigma$ uncertainty.
    In the bottom panel
    each bin contains the same number of sources, apart from the top bin which includes the
    strays (an additional two in the bottom plot only).
    These are centred on the mean values with the horizontal bar showing the range of
    the $Q_{\text{\HI}}$ bin and the vertical bar  $\pm1\sigma$ from the mean $\nu_{\rm TO}$.}
\label{Q_TO}
\end{figure}

\begin{figure}
 \includegraphics[scale=0.55]{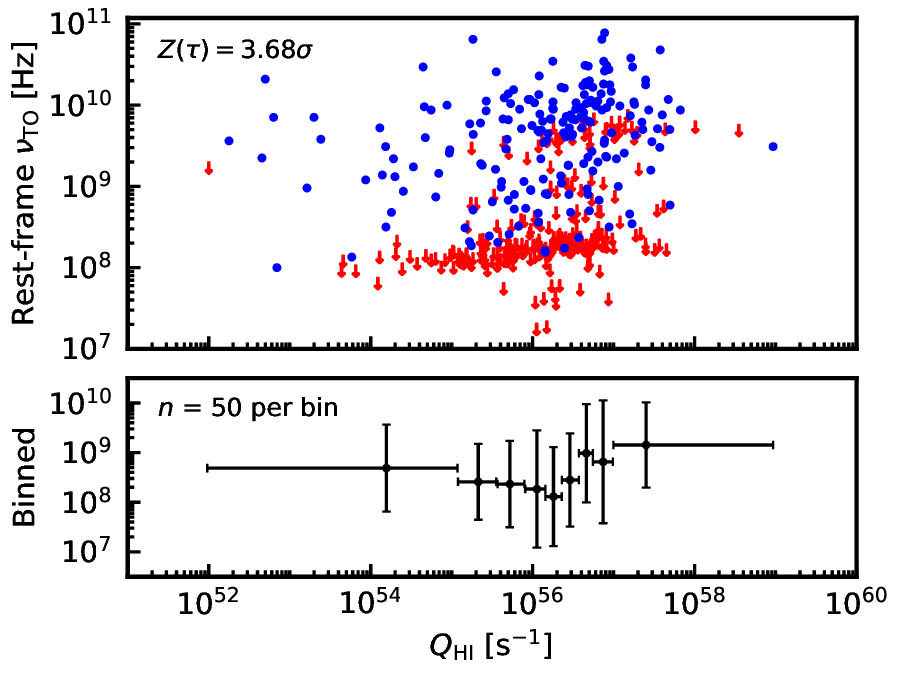}
   \vspace*{-0.6cm}
   \caption{As the bottom plot of Fig.~\ref{Q_TO}, but including the
     1st order fit sources as upper limits to the turnover frequency
     (shown as downwards arrows). For clarity, the error bars are not
     shown in the top panel. In the bottom panel the limits are
     included in the binned values via the \citet{km58}
     estimator.}
\label{Q_TO-limits}
\end{figure}

In Fig. \ref{Q_TO} we show the turnover frequency versus the ionising photon rate, 
from which we see a correlation between $\nu_{\rm TO}$ and
$Q_{\text{\HI}}$, which has a probability $p(\tau) \leq
9.10\times10^{-6}$ of arising by chance.  Given this correlation, we
now assume that the sources which exhibit no turnover have not been
searched to sufficiently low frequencies.  We include these sources by
assigning the lowest observed rest-frame frequency as the upper limit
to this (e.g. \citealt{ffs+90}). We also apply this to the 2nd order
fit sources where the turnover frequency yielded by the fit is lower
than the lowest observed rest-frame frequency (Fig.~\ref{Q_TO-limits}).

In order to incorporate the limits, we use the {\em Astronomy SURVival
  Analysis} ({\sc asurv}) package \citep{ifn86} which adds these as
censored data points, allowing a generalised non-parametric
Kendall-tau test.  Including these weakens the correlation ($p(\tau)
\leq2.3\times10^{-4}$)\footnote{For the uncorrected ionising photon
rates this is $p(\tau) = 1.63\times10^{-4}$ ($3.77\sigma$).}, which
may suggest, if $\nu_{\text{TO}}\propto Q_\text{\HI}$, that the 1st
order fit sources are not dominated by those which have not been
searched to low enough frequencies to detect the turnover.  However,
given the lower frequency ranges of the 1st order fit sources
(Sect.~\ref{rfl}), the detection of a turnover if the sampling were
ample cannot be ruled out. Hence, it is also possible that the
inclusion of these sources only acts to introduce noise.

\subsubsection{Caveats}

While we have found a strong correlation between the turnover frequency and the ionising photon
rate, one should be aware of:

\begin{enumerate}

\item For the one SDSS source which did match the low frequency catalogues,
  the lowest rest-frame frequency yielded by NED was 676~MHz, compared to
  141~MHz from the low frequency data (Sect.~\ref{paf}). Thus, until
  observed and included in NED, allowing automated photometry scraping,
  it is reasonable to assume that many of the SDSS (1st order fit) sample have not
  been observed to sufficiently low frequencies.
 
\item The 1st order fit sources  generally have sparser
sampling across the radio band (Sect.~\ref{rfl}). So, until the
power-law sources are sampled as comprehensively as those exhibiting a turnover,
we cannot be certain that a potential turnover does not occur within the
observed range.

\item As seen in Fig.~\ref{Q_TO}, there is a large variation in the uncertainties in the
  turnover frequencies. Two examples, at either end of the range, are shown in 
Fig.~\ref{lowest}.
\begin{figure}
 \centering\includegraphics[scale=0.44]{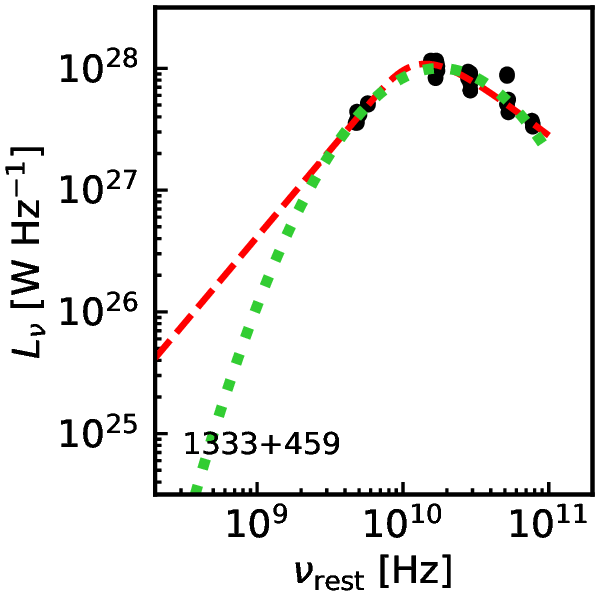}\hspace{3mm}
\centering\includegraphics[scale=0.44]{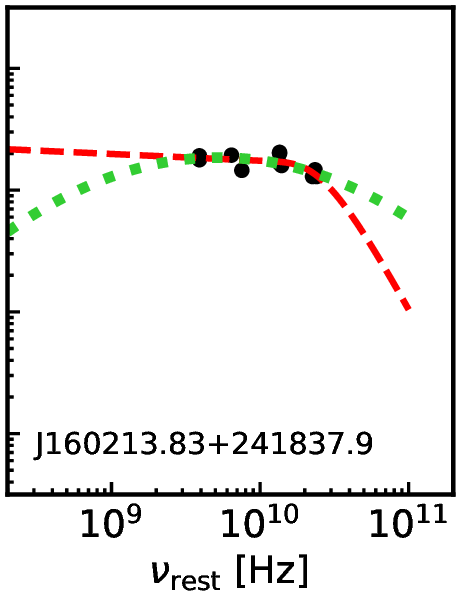}
\vspace*{-0.2cm}
\caption{Left: Example of a low uncertainty in the
  turnover frequency, with $\log_{10}\nu_{\rm{TO}} = 10.251\pm 0.002$ from 25 photometry points. Right:
   Example of a high uncertainty, with $\log_{10}\nu_{\rm{TO}} = 9.72\pm 1.86$ from nine photometry points.}
\label{lowest}
\end{figure}  
In a large enough sample we expect as may overestimates as
underestimates, so that the biases introduced by these would be
largely averaged out. Ideally, however, these should all be small and
so, again, more comprehensive sampling of the radio band is required.

\item From the upper limits in Fig.~\ref{Q_TO-limits}, we see that the
lowest observed frequency is correlated with the ionising photon rate,
which is most likely driven by the redshift, where sources are
observed to the same minimum frequency in a given survey,
confirmed in Fig.~\ref{freq_low}.
\begin{figure}
  \includegraphics[scale=0.53]{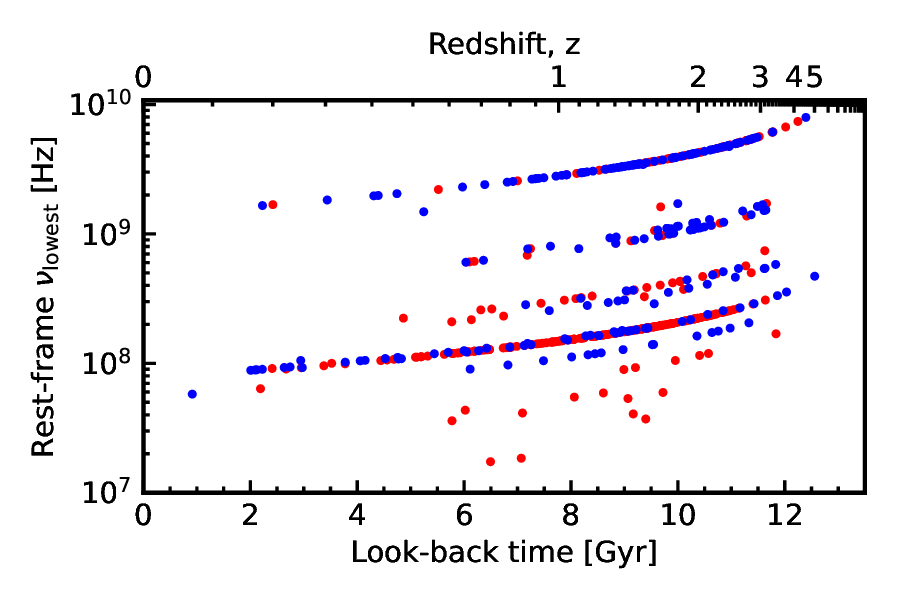}
  \vspace*{-0.3cm}
  \caption{The rest-frame minimum observed frequencies versus redshift.}
\label{freq_low}
\end{figure}
This is also evident in
\citet{gsi+23}, where the high redshifts ($z=5.6-6.6$) cause the low
observed frequencies ($\gapp0.1$~GHz) to be insensitive to turnover
frequencies below $\sim1$~GHz.
\end{enumerate}

\section{Modelling the \boldmath{$\nu_{\rm TO}-Q_{\text{\HI}}$} correlation}
\subsection{Dependence of the turnover frequency on the photo-ionisation rate}
\label{sec:dep}

We have found a correlation between the turnover frequency and the
ionising photon rate (Fig.~\ref{Q_TO}). Although this supports the hypothesis that there
is a connection between the turnover frequency and ionising photon rate
it is an empirical result. To explore the cause of the correlation
we now examine the theoretical dependence between the turnover frequency, as caused by free-free
absorption, and the number of photons available to ionise the neutral atomic gas.

For free-free  radiation the absorption coefficient is
\begin{equation}
  \kappa_{\nu} =  \frac{\sqrt{2\pi}}{32}\frac{Z^2e^6}{(4\pi\epsilon_0)^3}\frac{n_{\rm e}^2}{c(m_{\rm e}kT)^{3/2}\nu^2} \ln\left(\frac{3kT}{\nu h}\right),
\end{equation}
where $Z$ is the atomic number, $e$ the electronic charge, $n_{\rm e}$ the
electron density, $k$ the Boltzmann constant, $T$ the temperature, $\epsilon_0$ the
permittivity of free space, $c$ the speed of light and $m_{\rm e}$ the electron mass.

For neutral hydrogen ($Z=1$), this can be written as 
\begin{equation}
  \kappa_{\nu} = 0.0719\left(\frac{n_{\text{e}}}{\nu}\right)^2\frac{1}{T^{3/2}}\ln\left(\frac{3kT}{\nu h}\right),
\label{kappa2}
\end{equation} 
per metre, where $n_e$ is in \ccm.

For a spherical \HII\ region of radius $r_{\rm str}$, at a luminosity distance $D_{\text{L}}$, the flux density is
\begin{equation}
  S_{\nu} = (1-e^{-\tau_{\nu}}) \pi\left(\frac{r_{\rm em}}{D_{\text{L}}}\right)^2B_{\nu},
  \label{eqS}
\end{equation}
where $r_{\rm em} = r_{\rm str}$ for a sphere of constant density,
the optical depth is $\tau_{\nu} = \int\kappa_{\nu}dr$
and $B_{\nu}$ is the specific brightness, which, in the Rayleigh-Jeans regime ($h\nu\ll kT$)
is given by
\begin{equation}
  B_{\nu}\approx \frac{2\nu^2}{c^2}kT.
  \label{RJ}
\end{equation}
In equilibrium, the rate of ionisation is balanced by the rate of recombination \citep{ost89}, 
\begin{equation}
 Q_{\text{\HI}}\equiv \int^{\infty}_{\nu_{\rm ion}}\frac{L_{\nu}}{h\nu}\,d{\nu}= 4\pi\int^{r_{\rm str}}_{0}\,n_{\rm p}\,n_{\rm e}\,\alpha_{A}\,r^2\, dr,
\label{Q1}
\end{equation}
where $n_{\rm p}$ is the proton density and 
$\alpha_{A}$ the radiative recombination rate coefficient of hydrogen ($4.19\times10^{-13}$~cm$^3$~s$^{-1}$
at the canonical $T=10\,000$~K,
\citealt{of06}).\footnote{http://amdpp.phys.strath.ac.uk/tamoc/DATA/RR/}

\subsubsection{Constant density profile}

\begin{figure}
  \includegraphics[scale=0.48]{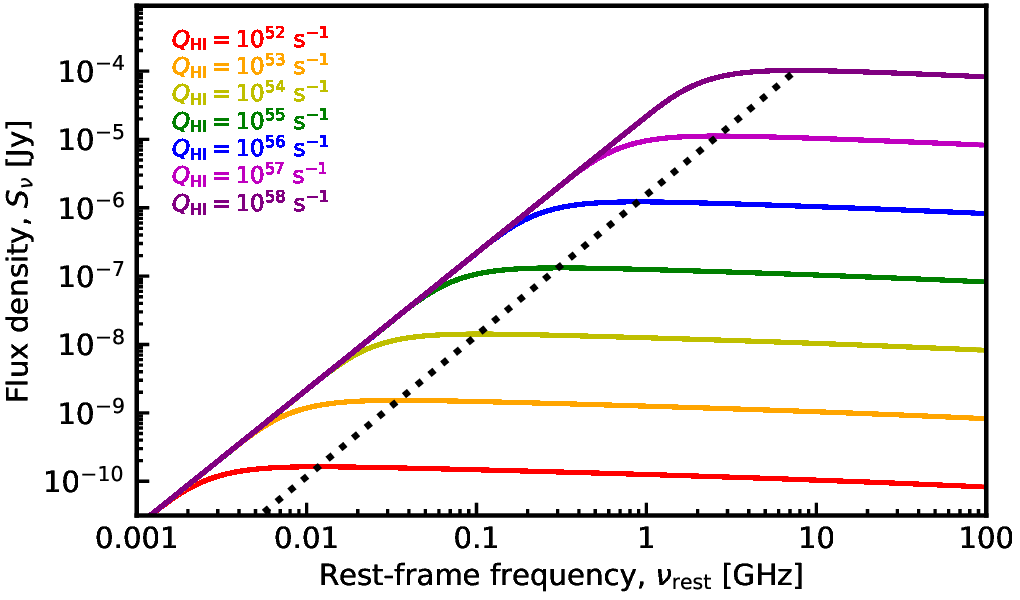}
  \vspace{-2mm}
  \caption{The radio SEDs for different photo-ionising rates for a
    $r_{\rm str}=1$~kpc sphere of $T=10\,000$~K gas of constant density at $z=1$. The
    flux density is for $r_{\rm em} = r_{\rm str}$.
    The densities which can be completely ionised within this sphere range from $n=0.44$~\ccm\
    for $Q_{\text{\HI}} = 1\times10^{52}$~s$^{-1}$ to $n=440$~\ccm\ for
    $Q_{\text{\HI}} = 1\times10^{58}$~s$^{-1}$. The black dotted line shows the increase of $\nu_{\text{TO}}$
    (as given by the maximum flux density) with $Q_{\text{\HI}}$.} 
\label{S_nu-Q}
\end{figure}

From Equ.~\ref{Q1}, the ionising photon rate required to completely
ionise a neutral plasma ($n_{\rm p} = n_{\rm e}$) of
constant density within a Str\"{o}mgren sphere of radius $r_{\text{str}}$ is
\begin{equation}
Q_{\text{\HI}} = \frac{4}{3}\pi \alpha_{A}n_{\rm e}^2 r_{\text{str}}^3.
\end{equation}
For a range of ionising photon rates we use this to obtain the electron density,
which we insert into Equ.~\ref{kappa2}, giving the absorption
coefficient $\kappa_{\nu}$.  For a constant electron density through a
path $r_{\text{str}}$, $\tau_{\nu} = \kappa_{\nu}r_{\text{str}}$
which, via Equ.~\ref{eqS}, gives the variation in flux density with
frequency (Fig.~\ref{S_nu-Q}).
From this, it is clear that for a given density, temperature and radius,
the turnover frequency is correlated
with the ionising photon rate, as is seen for the observational data (Sect.~\ref{sec:res}).

\subsubsection{Exponential density profile}
\label{edp}

A more physically realistic model, especially on galactic scales, is
an exponential decrease in gas density with distance from the nucleus
of the galaxy (e.g. \citealt{bal82,bbs91}).
\citet{ob97} use a distribution of the form $n =n_0 r^{-a}$, where $0\lapp a \lapp3$.
  However, this gives $n = 0$ at $r = 0$ and an infinite column density for
$N_{\text{\HI}} = \int_0^{\infty} n dr$.
  We therefore use the exponential sphere model, $n = n_0e^{-r/R}$,
  which gives $N_{\text{\HI}} = \int_0^{\infty} n dr = n_0R$.

\begin{figure}
  \includegraphics[scale=0.55]{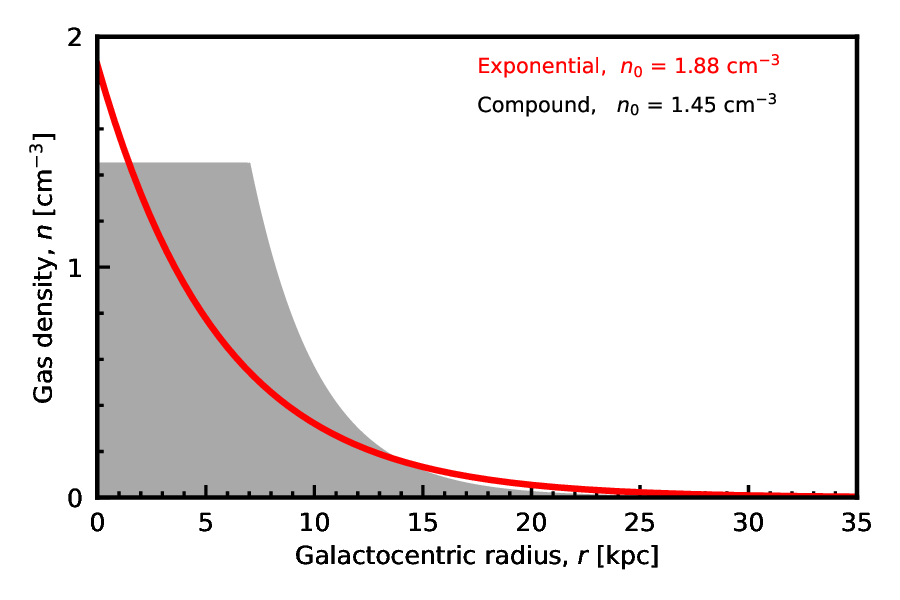}
  \vspace*{-0.4cm}
  \caption{The gas density versus the galactocentric radius for the
    simple exponential (red) and compound (grey, see Sect.~\ref{sec:comp}) models of the Milky Way.}
\label{n_dist}
\end{figure}

\begin{figure*}
   \centering\includegraphics[scale=0.5]{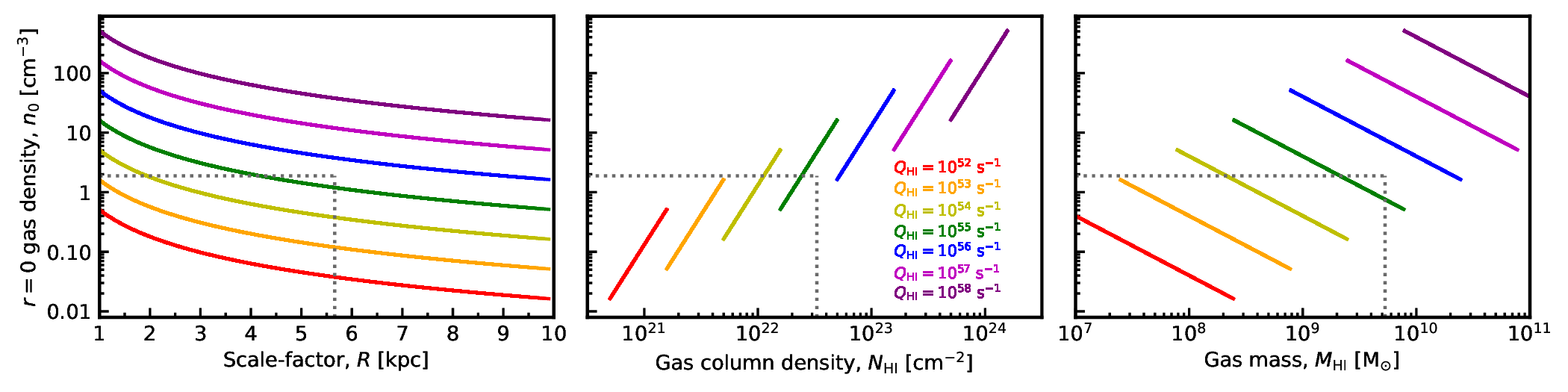}
 \vspace*{-0.7cm}
 \caption{Left: The central gas density--scale-length degeneracy for complete ionisation
   of an exponential gas distribution as constrained by Equ.~\ref{Q2}. Middle: The resulting column densities from $N_{\text{\HI}} = n_0R$. Right: The resulting gas masses, assuming the Galactic flare factor \citep{kk09}. The dotted line shows the values for the Galactic exponential distribution (Fig.~\ref{n_dist}), where
 all of the gas is ionised for $Q_{\text{\HI}} > 2.5\times10^{55}$~s$^{-1}$.}
 \label{3_Q}
\end{figure*}

\begin{figure*}
  \includegraphics[scale=0.53]{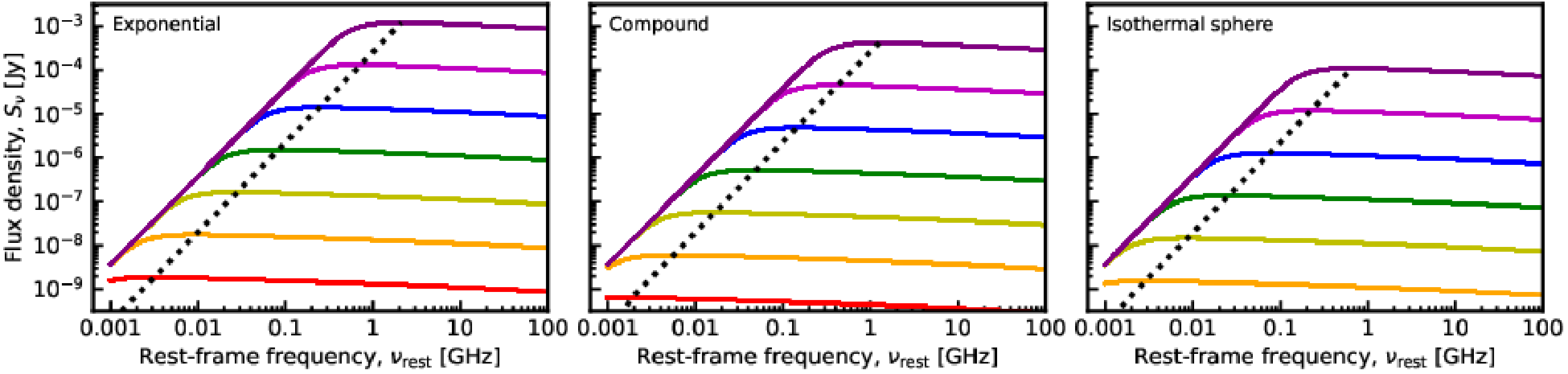}
  \vspace{-2mm}
  \caption{As Fig.~\ref{S_nu-Q}. Left: For an exponential density
    profile. The central densities for which the gas can be completely ionised
    within this sphere range from $n_0=0.04$~\ccm\ for $Q_{\text{\HI}} = 10^{52}$~s$^{-1}$ to $n_0=37$~\ccm\ for $Q_{\text{\HI}} = 10^{58}$~s$^{-1}$.
    Middle: For a compound density
    profile, using the Galactic values of $r_0 = 7$~kpc and $R=3.15$~kpc.
    The central densities for which the gas can be completely ionised
    within this sphere range from $n_0=0.02$~\ccm\
    for $Q_{\text{\HI}} = 1\times10^{52}$~s$^{-1}$ to $n_0=16.6$~\ccm\
    for $Q_{\text{\HI}} = 1\times10^{58}$~s$^{-1}$.
    Right: For a spherical density
    profile. The central densities for which the gas can be completely ionised
    within this sphere range from $n_0=0.003$~\ccm\ for $Q_{\text{\HI}} = 10^{52}$~s$^{-1}$ to $n_0=33$~\ccm\ for $Q_{\text{\HI}} = 10^{58}$~s$^{-1}$. } 
\label{S_nu-Q_exp}
\end{figure*}

  Using the Milky Way's $ N_{\text{\HI}}= 3.3\times10^{22}$~\scm\ and
  breaking the  $n_0R$ degeneracy via the gas mass \citep{cw12} 
\begin{equation}
  M_{\text{gas}} = \frac{2\pi m_{\rm p} n_0}{F}\int_0^{\infty}e^{-r/R}r^2dr = \frac{4\pi m_{\rm p} n_0R^3}{F},
\label{mass_eq}
\end{equation}
  where $F = 20$ is the Galactic flare factor \citep{kk09}, gives $M_{\text{gas}} =5.3\times10^{9}$~\Mo, $n_0 = 1.88$~\ccm\ and $R=5.66$~kpc for the Milky Way (Fig.~\ref{n_dist}).
 For the exponential sphere complete ionisation of the gas occurs when the ionising photon rate reaches \citep{cw12}
\begin{equation}
  Q_{\text{\HI}} = \pi \alpha_{A}n_0^2R^3.
  \label{Q2}
\end{equation}
Using Equ.~\ref{Q2} to obtain $n_0$ for a given $R$, where the gas is completely ionised (Fig.~\ref{3_Q}),
gives the optical depth through the gas as (Equ.~\ref{kappa2})
\[
  \tau_{\nu} = 0.0719\left(\frac{1}{\nu}\right)^2\frac{1}{T^{3/2}}\ln\left(\frac{3kT}{\nu h}\right)n_0^2\int_0^{\infty} e^{-2r/R}dr.
  \] 
Integrating this gives
\begin{equation}
\tau_{\nu} = \frac{0.0719}{2}\left(\frac{1}{\nu}\right)^2\frac{1}{T^{3/2}}\ln\left(\frac{3kT}{\nu h}\right)n_0^2 R,
\label{kappa3}
\end{equation}
which, again via Equ.~\ref{eqS}, gives the variation in flux density
with frequency (Fig.~\ref{S_nu-Q_exp}, left). Like the constant density
distribution, this exhibits an increase in the turnover frequency with
ionising photon rate.

Although, unlike the constant density model, at large distances the path
length is not required for the
optical depth calculation,
a value is required for the flux density (Equ.~\ref{eqS})
and we have assumed that $r_{\rm em} = 2.3R$ as the extent over which the
emission can be detected. However, while the flux densities are not absolute, the
turnover frequency is independent of the choice of $r_{\rm em}$.

\subsubsection{Constant and exponential density (compound) profile}
\label{sec:comp}

For the Milky Way the gas distribution is a combination of both the constant and
exponential profile \citep{kk09};
\[ 
n = 
\left\{
\begin{array}{l l } 
n_0, &  r \leq r_0\\
n_0 e^{-(r-r_0)/R},&  r > r_0,
\end{array} \right. 
\]
where $n_0 = 1.45$~\ccm, $R=3.15$~kpc and $r_0 = 7$~kpc (see Fig.~\ref{n_dist}).
For this distribution complete ionisation of the
gas occurs when the ionising photon rate is \citep{cur24}
\begin{equation}
Q_{\text{\HI}}= \pi \alpha_{A} n_0^2 \left(\frac{4r_0^3}{3} + R\left[2r_0^2 + 2Rr_0 + R^2\right]\right).
\label{Q_comp}
\end{equation}
Again, assuming that all of the gas is ionised, we use the electron density for a given ionising
photon rate (Equ.~\ref{Q_comp}) to calculate
the optical depth through the gas
\[
\tau_{\nu} = 0.0719\left(\frac{1}{\nu}\right)^2\frac{1}{T^{3/2}}\ln\left(\frac{3kT}{\nu h}\right)n_0^2\left(\int_0^{r_0}dr + \int_{r_0}^{\infty} e^{-2r/R}dr\right),
\]
which gives
\begin{equation}
\tau_{\nu} = \frac{0.0719}{2}\left(\frac{1}{\nu}\right)^2\frac{1}{T^{3/2}}\ln\left(\frac{3kT}{\nu h}\right)n_0^2\left(r_0 + \frac{R}{2}\right).
\label{kappa4}
\end{equation}
This results in the frequency profiles shown in Fig.~\ref{S_nu-Q_exp} (middle), 
where, again, we have assumed $r_{\rm em} = 2.3R$ as the size of the emission region,
although this does not affect the apparent $Q_{\text{\HI}}-\nu_{\rm TO}$ correlation.

\subsubsection{Elliptical profiles}
\label{ell}

\begin{figure}
  \includegraphics[scale=0.55]{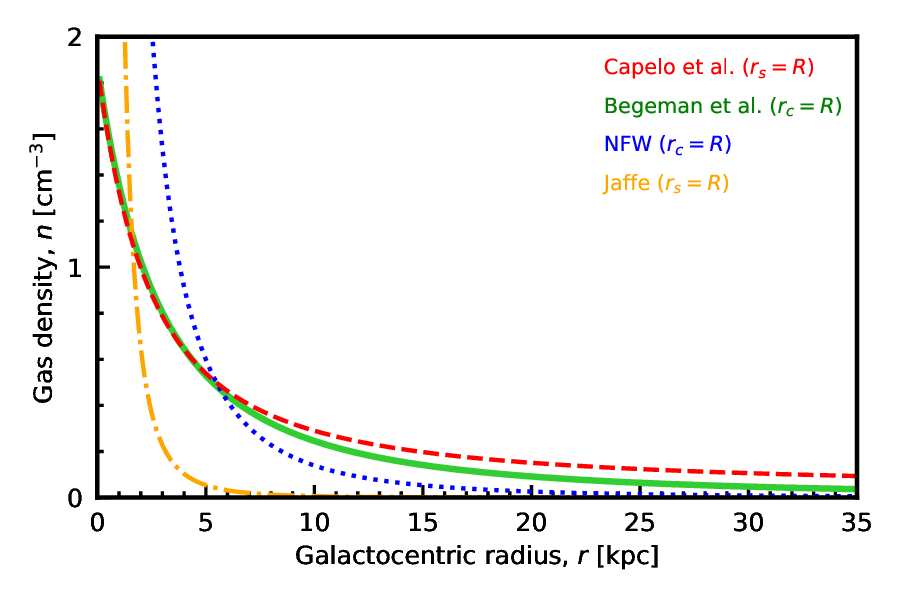}
  \vspace*{-0.6cm}
  \caption{The gas density versus the galactocentric radius for the \citet{jaf83},
    \citet{bbs91}, \citet{nfw96} and \citet{cnc10} models. In all cases we use the
    scale-length and central density of the simple exponential model of the
    of the Milky Way (Fig.~\ref{n_dist}).}
\label{n_DM}
\end{figure}

Since our sample comprises radio loud AGN, it is likely that most of
the host galaxies are elliptical. These can be modelled via several
similar density profiles, e.g.  the Jaffe profile \citep{jaf83}, which
models the distribution of light in a spherical galaxy as $n =
n_0\,(r_{\rm s}/r)^2/4\,\pi\,(1 + r/r_{\rm s})^2$, where $r_{\rm s}$
is the radius which contains half the total emitted light. However, as
seen in Fig.~\ref{n_DM} this has an asymptotic density distribution,
as does the NFW profile \citep{nfw96}, $n = n_0\,(r_{\rm c}/r)/(1 +
r/r_{\rm c})^2$, where $r_{\text{ c}}$ is the radius of the halo core.

These can be avoided by employing the halo density distribution of an
isothermal sphere, for example  $n = n_0/(1 + r/r_{\text{ c}})^2$
\citep{bbs91} or equation 7 of \citet{cnc10},
\[
n = n_0\exp\left[-4\pi G\delta_{\text{c}}\rho_{\text{c}}r_{\text{s}}^2\frac{\mu m_{\text{p}}}{kT} \left(1 - \frac{\ln(1 + r/r_{\text{s}})}{r/r_{\text{s}}}\right)\right], 
\]
where $G$ is the gravitational constant, $\rho_{\text{c}}$ is the
critical density of the Universe $r_{\text {s}}$ is a scale radius,
$\mu$ the mean molecular weight of the gas and $m_{\text{p}}$ the
proton mass. We select this, the simplest model, cf. the polytropic
and stellar models, to limit the number of free parameters. Using the column
density of the Milky Way at $r = 35$~kpc ($N_{\text{\HI}} =3.25\times10^{22}$~\scm)
to constrain the characteristic density gives $\delta_{\text{c}} = 2370$ for 
 $T=10\,000$~K and $r_{\text{s}} = R$.

Given its relative simplicity, while yielding a similar density profile to that
of \citet{cnc10}, we use the profile of \citeauthor{bbs91}, for which complete ionisation
of the gas occurs when 
\[
Q_{\text{\HI}} = \frac{4}{3}\pi \alpha_{A}n_{\rm e}^2 r_{\text{c}}^3.
\]
Proceeding as above, the  optical
depth through the completely ionised gas
is
\[
 \tau_{\nu} = 0.0719\left(\frac{1}{\nu}\right)^2\frac{1}{T^{3/2}}\ln\left(\frac{3kT}{\nu h}\right)n_0^2\int_0^{\infty} \frac{dr}{(1 + r/r_{\text{c}})^4},
\]
giving
\begin{equation}
\tau_{\nu} = \frac{0.0719}{2}\left(\frac{1}{\nu}\right)^2\frac{1}{T^{3/2}}\ln\left(\frac{3kT}{\nu h}\right)n_0^2 \frac{r_{\text{c}}}{3}.
\label{kappa_DM}
\end{equation}
which has a similar form to the exponential density (Equ.~\ref{kappa3}) and 
frequency (Fig.~\ref{S_nu-Q_exp}, right) profiles.

\subsubsection{Summary}

For all of the above models $\nu_{\rm T0}\propto
Q_{\text{\HI}}^{~~0.97}$, although there is some variation in the
constant of proportionality and each model therefore predicts quite
different values for the turnover frequency,
\begin{table}
  \centering
  \begin{minipage}{80mm}
    \caption{The turnover frequencies predicted from the different gas models for different
      ionising photons rates. The last column is for the isothermal sphere in Sect.~\ref{ell}.}
\begin{tabular}{@{}l r r r r  @{}} 
\hline
\smallskip
$Q_\text{\HI}$ [s$^{-1}$] & \multicolumn{4}{c}{Gas density model}\\
 &   Constant & Exponential & Compound & Iso. Sphere\\
\hline
$10^{52}$          & 11~MHz   &    1.9~MHz &  8.5~MHz & 1.0~MHz\\
$10^{53}$          & 35~MHz       &     5.6~MHz &  25~MHz & 2.9~MHz\\
$10^{54}$          & 100~MHz    &     16~MHz &  74~MHz & 8.7~MHz\\
$10^{55}$          & 300~MHz  &     49~MHz & 220~MHz & 26~MHz\\
$10^{56}$          & 0.89~GHz &     140~MHz & 0.65~GHz & 76~MHz\\
$10^{57}$          & 2.6 GHz &     0.43~GHz & 1.9~GHz & 0.22~GHz\\
$10^{58}$          & 8.8~GHz   &   1.3~GHz & 5.5~GHz & 0.66~GHz\\
\hline  
\end{tabular}
\label{T1}  
\end{minipage}
\end{table}
which can vary by up to an order of magnitude for a given ionising
rate (Table~\ref{T1}). Only where there is a constant density
component, will we see a turnover at $\nu_{\rm TO}\sim1$~GHz for
$Q_{\text{\HI}}\sim10^{56}$~s$^{-1}$ (sufficient to ionise all of the
gas in a large spiral). For the exponential and spherical models it
appears that $Q_{\text{\HI}}\gapp10^{58}$~s$^{-1}$ is required for $\nu_{\rm TO}\sim1$~GHz,
which is consistent with the hypothesis that sources with no
turnover are not subject to complete ionisation. 

\subsection{The turnover frequency of J0002+2550}
\label{TOJ0002+2550}

Equipped with the expected value of the turnover frequency for a given photo-ionisation rate,
we can estimate the turnover for J0002+2550 (Sect.~\ref{sect:J0002+2550}).
At $z = 5.80$, the luminosity distance to the source is $D_{\rm L} =56\,900$~Mpc
and the finest beam size of $3.4''$, in which the source is
unresolved \citep{gsi+23}, gives $r_{\rm em}\lapp10$~kpc for the
low frequency emission.

\begin{figure}
  \includegraphics[scale=0.55]{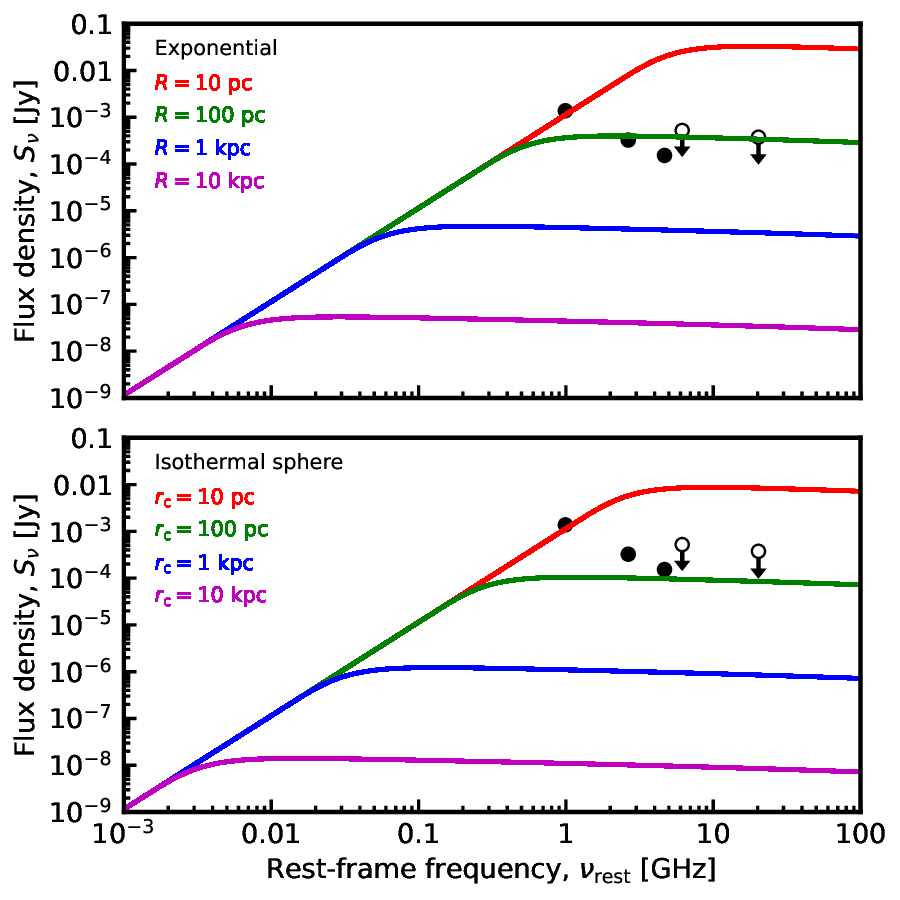}
   \vspace*{-0.4cm}
   \caption{The radio SED from an $r_{\rm em} = 10$~kpc \HII\ region
     at $z = 5.80$ due to a $Q_{\text{\HI}} =
     1.0\times10^{55}$~s$^{-1}$ ionising rate for different
     scale-lengths of the gas density distribution.  The points and
     upper limits show the photometry from \citet{gsi+23}.}
  \label{S_nu-gsi}
\end{figure}

Again, assuming complete ionisation by the observed $Q_{\text{\HI}} = 9.3\times10^{54}$~s$^{-1}$
(Sect.~\ref{sect:J0002+2550}), in 
Fig.~\ref{S_nu-gsi} we show the expected radio SED for various scale-lengths
from which it is apparent that, for a scale-length of $R\gapp0.1$~kpc, the turnover frequency is
$\lapp1$~GHz.
Although consistent with the absence of a turnover at $\nu_{\rm TO}\gapp1$~GHz,
our simple free-free radiation model (which forces  $\alpha_{\rm thick} \approx2$ and
$\alpha_{\rm thin} \approx-0.1$) does not trace the steep spectral index of the
optically thin portion of the \citeauthor{gsi+23} spectrum
(where $\alpha_{\rm thin} \approx-1.6$), which they attribute to the ionised
medium producing synchrotron emission in addition to a spectral break 
caused by ageing electrons.

We note that the predicted flux density approaches the observed
$\sim1$~mJy at $\nu_{\text{rest}}\sim1$~GHz, for a scale length of
$R\sim100$~pc. This is about an order of magnitude lower than that
of the Galaxy 
but, due to hierarchical build-up, we expect smaller galaxies
\citep{bmce00,lf03} and, thus, smaller scale-lengths at high
redshift. However, we have set the extent of the emission to the minor
axis of the smallest beam in which the source is unresolved, and so there is some
wriggle room in the absolute flux measurements.

Correcting the  UV spectral index for J0002+2550 gives $\alpha_{\rm UV} = -9.4$,
which results in an ionising photon rate of $Q_\text{\HI} =  1.0\times10^{55}$~s$^{-1}$,
which changes little from the uncorrected value. From Table~\ref{T1}, we see that
the exponential and spherical gas density profiles allow for no turnover at $\nu\gapp~1$~GHz for
relatively high ionising rates ($Q_\text{\HI}\lapp10^{58}$~s$^{-1}$), whereas
the constant density model precludes this at $Q_\text{\HI}\gapp10^{56}$~s$^{-1}$.
Unfortunately, the other targets of \citeauthor{gsi+23} have
insufficient rest-frame UV photometry to see if such steep UV spectra apply to the
sample as a whole (see Appendix C). As it is, this source is consistent with a low
level of ionisation causing shifting the turnover frequency to below the minimum
observed value, but with the caveat that the photometry correction for intervening absorption is
statistical only.

\subsection{Turnover frequency and size}

\subsubsection{Observations}

\begin{figure}
 \hspace*{-0.1cm}
  \includegraphics[scale=0.53]{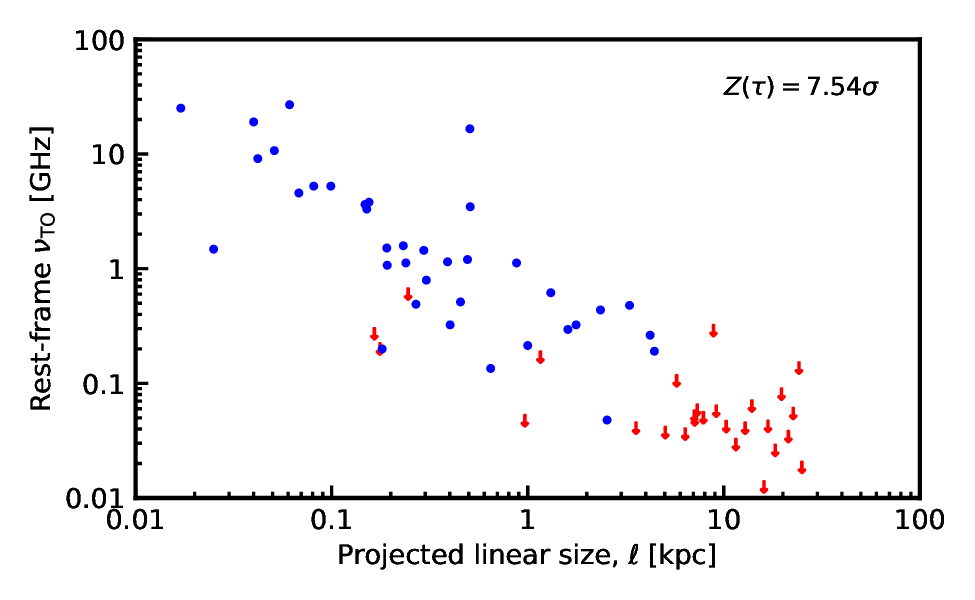}
   \vspace*{-0.5cm}
   \caption{The turnover frequencies versus the projected linear size for the sample of \citet{ob97}.}
\label{ob97_LS-TO}
\end{figure}

As stated in Sect.~\ref{intro}, the turnover frequency is
anti-correlated with the projected linear size, which is clearly
evident in the \citet{ob97} data
(Fig.~\ref{ob97_LS-TO}).\footnote{$\ell$ is calculated using updated
spectroscopic redshifts and cosmological parameters \citep{paa+20}.}
On average, the turnover frequencies we derive are slightly lower than
those of \citeauthor{ob97}, which could be due to the recent increase in the 
availability of low frequency data (see Appendix B). Out of the 66
sources with a measured linear size, 38 exhibit a turnover, 27 are flagged as upper limits, since the fit
gives a turnover below the lowest observed frequency,
and one has an inverted spectrum (FBQS\,J133037.6+250910).
Including the limits, and excluding J1330+2509, we obtain $p(\tau)=4.56\times10^{-14}$ for the
correlation, which is similar to that using the values
of \citeauthor{ob97} directly, where there are no
limits.\footnote{$p(\tau) = 1.58\times10^{-13}$ ($7.38\sigma$).}

\subsubsection{Model}

If the extent of the radio emission scales with $R$, the
anti-correlation between turnover frequency and size is evident from
our model. Furthermore, for an ionising photon rate of $Q_{\text{\HI}}
\sim10^{55}$~s$^{-1}$ (J0002+2550, Fig.~\ref{S_nu-gsi}) and
$R\sim\ell$, the model is in agreement with the observations
(Sect.~\ref{intro}); that is, $\nu_{\rm TO} \sim0.4 - 5$~GHz for
$R\lapp1$~kpc (HFPs/GPSs) and $\nu_{\rm TO}\lapp0.5$~GHz for
$R\gapp10$~kpc (CSSs).

In order to generalise this, we need to break the degeneracy between
$n_0$ and $R$ in Equ.~\ref{kappa3}.  For this we use the exponential
model for which we can use the fact that, for a given mass, the
product $n_0 R^3$ is constant (Equ.~\ref{mass_eq}).\footnote{For the
spherical model (Sect.~\ref{ell}), $M_{\text{gas}} = 2\pi m_{\rm p}
n_0 \int_0^{\infty} r^2dr/(1+r/r_{\text{c}}^2) = 4\pi\,n_0\,r_{\rm
  c}^3[1 + \frac{r}{r_{\rm c}} - \frac{r_{\rm c}}{r_{\rm c} + r} -
  2\ln(r_{\rm c} + r) + 2\ln(r_{\rm c})]_0^{\infty}$.}  From a
low-redshift survey of 21-cm emission from the 1000 \HI\ brightest
galaxies in the southern sky, \citet{ksk+04} find a range of $M_{\rm
  gas} \sim10^7 - 10^{11}$\,\Mo, with a median of $M_{\rm gas}
=2.9\times10^9$\,\Mo.  Using these masses and assuming the Galactic
flare factor, for a given scale-length we determine $n_0$, which is
then used in Equs.~\ref{kappa3} and \ref{eqS} to produce an SED (as
for Fig.~\ref{S_nu-gsi}), from which we obtain the turnover
frequency.\footnote{In Equ.~\ref{eqS}, $r_{\rm em}$ and $D_{\rm L}$
are only required to obtain the absolute flux density.}

\begin{figure}
  \includegraphics[scale=0.53]{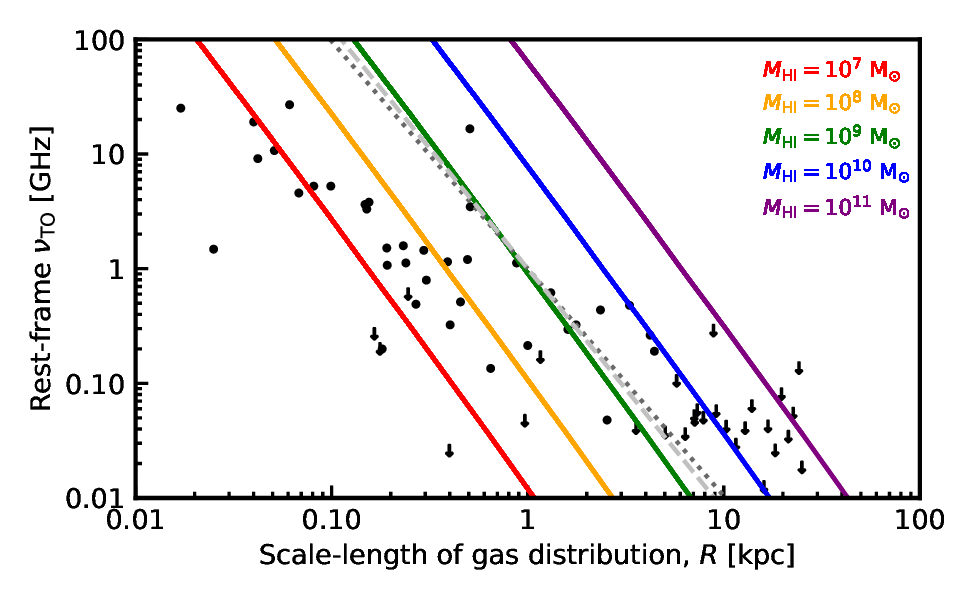}
   \vspace*{-0.6cm}
   \caption{The turnover frequency at various total gas masses for completely ionised gas of
     scale-length $R$. The markers show the data of \citeauthor{ob97} (Fig.~\ref{ob97_LS-TO}), assuming
     $\ell = R$ 
     and the dotted and dashed lines their models 1 \& 2, respectively.}
     \label{TO_L-mass}
\end{figure}
The results are shown in Fig.~\ref{TO_L-mass}, where we see the
predictions overlap the observations of \citet{ob97} for the range of
gas masses expected (assuming $R = \ell$).  We also note that the
median gas mass of $M_{\rm gas} =2.9\times10^9$\,\Mo\ passes close to
the middle of the distribution of the observed data. This is also
closely aligned with the two models of \citeauthor{ob97}, which are based
on synchrotron self-absorption models.
Thus, for a gas scale-length
close to the radio source size, our simple free-free model
can reproduce the observed $\nu_{\rm TO}$--$\ell$ relationship. 

\begin{figure}
  \includegraphics[scale=0.57]{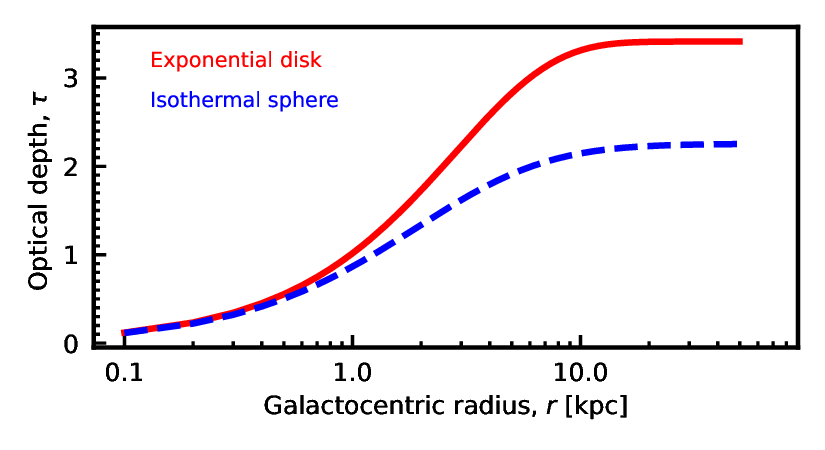}
   \vspace*{-0.2cm}
   \caption{The optical depth of the 100~MHz absorption versus the distance traversed through the
     gas for the  $R = 5.66$~kpc and $n_0 = 1.88$~\ccm\ exponential density profile (Sect.~\ref{edp})
     and the  $r_{\text{c}} = 5.66$~kpc and $n_0 = 1.88$~\ccm\ spherical density profile (Sect.~\ref{ell}).}
   \label{tau_r}
   \end{figure}

Lastly, we note that, while we are considering the ionisation of the large-scale gas,
the material causing the turnover in the radio SED may be mainly located within the
central $\sim1$~kpc (e.g. \citealt{dscf00,cge+15,mmo+18}). However, for the exponential
and isothermal sphere models of the gas density,
we see that at 100~MHz the absorption is already optically thick  within this region (Fig.~\ref{tau_r}). 

\subsection{Ionising photon rates of the \citeauthor{ob97} sample}

Of the \citeauthor{ob97} sample, only 11 sources have sufficient rest-frame photometry
to yield an ionising photon rate, comprising five with a turnover within the sampled
frequency range and six without.
\begin{figure}
  \includegraphics[scale=0.53]{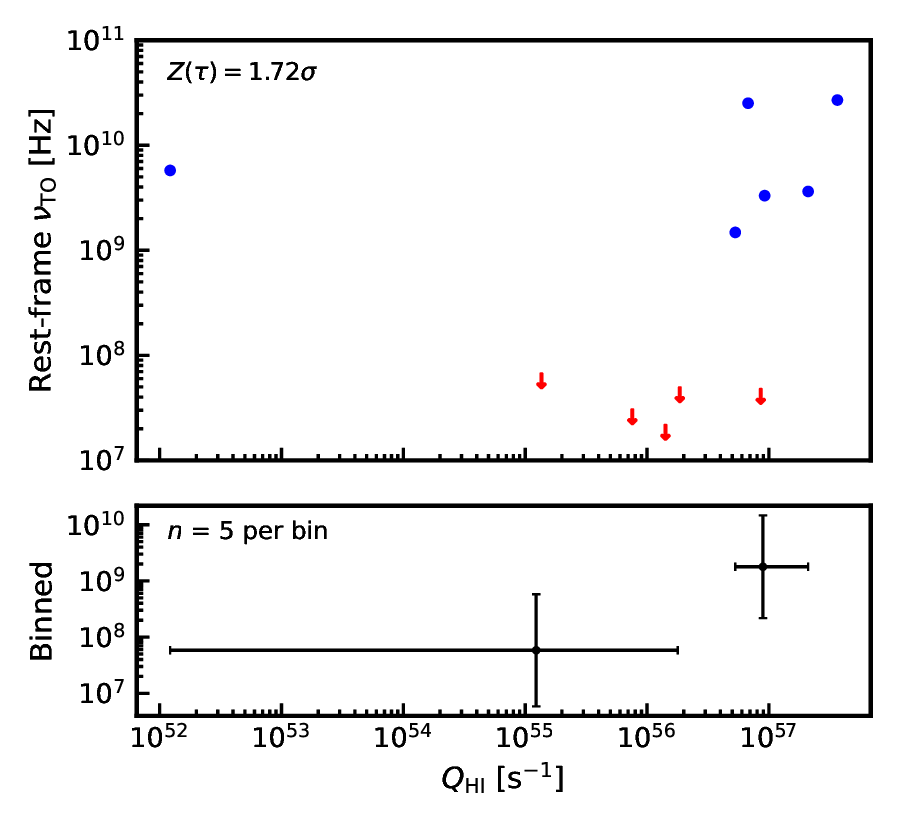}
   \vspace*{-0.4cm}
   \caption{The turnover frequency versus the ionising photon rate for the sample of \citet{ob97}.}
        \label{ob97-Q_TO}
\end{figure}
As seen from Fig.~\ref{ob97-Q_TO}, five detections and six limits are insufficient to yield a convincing
correlation ($p(\tau) = 0.085$), although one is suggested by the binning.

\section{Discussion and conclusions}

It has long been known that the turnover frequency in the SEDs of
extragalactic radio sources is anti-correlated with the projected size
of the emission. If the profile of the radio SED is dominated by
free-free absorption we way may expect the turnover frequency,
$\nu_{\rm TO}$, to be related to the rate of ionising photons emitted by the
AGN. Thus, we suggest that
radio sources which do not exhibit a turnover have low ionising photon rates,
causing the turnover to be below the minimum observed frequency.

To test this, we initially examined the observational data, mining all of the radio, IR, optical and
UV photometry, in order to obtain the turnover frequency and 
ionising photon rate:
\begin{itemize}
  
\item For the sample of \citet{gsi+23}, where only one of the nine high redshift sources
  shows any indication of a turnover:
  Sufficient rest-frame UV photometry was available for only one
  source, J0002+2550. This was found to have a relatively
  low ionising photon rate of
  $Q_{\text{\HI}}\sim 10^{55}$~s$^{-1}$, which is consistent
  with a low turnover frequency.

\item For a larger sample, optically selected in order to increase the number of sources
  with comprehensive UV photometry:
\begin{itemize}
\item For the sources which exhibit a turnover in their radio SED, we find a
  correlation between  $\nu_{\rm TO}$ and $Q_{\text{\HI}}$, significant at $\approx4.2\sigma$.
\item By assuming that the radio sources fit by a power-law have their
  turnover frequencies below the minimum observed, the 
  significance reduces to $\approx3.7\sigma$. However, there is strong evidence
  that the power-law sources are not as comprehensively sampled in the radio band than
  those exhibiting a turnover, and, if these do have a putative undetected peak in the spectrum,
  including these adds noise to the data. Thus, more complete sampling in the radio band for  the power-law sources
  is required.
\end{itemize}
\end{itemize}
While compelling, the correlation between the turnover frequency and the ionising photon rate
is an empirical result and so, to provide a physical justification, we
produced a model relating $\nu_{\rm TO}$ to  $Q_{\text{\HI}}$, via the electron density.
This is done by assuming purely free-free radiation from a completely ionised galactic disk/sphere.
From this we find;

\begin{itemize}

  \item For gas of a constant density, an exponential density profile,
    a combination of both (as per the Milky Way) and a spherical
    density profile, that the turnover frequency is correlated with
    ionising photon rate, which must exceed that to ionise all of the
    gas in a large spiral ($Q_{\text{\HI}}\gapp10^{56}$~s$^{-1}$) in
    order to exhibit $\nu_{\rm TO}\gapp1$~GHz. This supports the
    hypothesis that sources with low ionising photon rates
    ($Q_{\text{\HI}}\ll 10^{56}$~s$^{-1}$) have their turnover
    frequencies below the minimum observed ($\nu_{\rm TO}\ll 1$~GHz)

\item That the exponential and spherical density models can fit the SED of J0002+2550 for
  $R\gapp100$~pc, suggesting that the turnover frequency is
  indeed below the lowest observed rest-frame frequency of 1~GHz.

\item For a given mass of neutral gas, and using the scale-length as a
  proxy for the projected linear size, the dependence of turnover
  frequency on the scale-length of the exponential gas distribution
  closely traces the observational data and the synchrotron
  self-absorption model.
\end{itemize}
The model, which relies solely on ionisation of the atomic gas by the
AGN, is considerably simpler than the synchrotron self-absorption
models \citep{bdo97,ob97}, which have to assume a large number of  magnetic
field reversals to keep net polarisation at the low values which are observed.

However, given the sparseness of the radio data, when prioritising
sources with measured UV luminosities, we cannot reliably determine
spectral indices below and above the turnover. A further caveat is we
assume that the gas is completely ionised, although the fact that all
of the distributions tested give the same dependence $\nu_{\rm
  T0}\propto Q_{\text{\HI}}^{~~0.97}$ may suggest that the model
remains valid over a region of ionisation enclosed within an envelope
of neutral gas.  We also assume that the scale-length of the gas
distribution is proportional to the extent the emission. This is a
reasonable assumption and only affects the flux density of the
modelled source and not the turnover frequency.  Interestingly,
setting the scale-length $R = \ell$ does reproduce the $\nu_{\rm
  TO}$--$\ell$ synchrotron self-absorption models \citep{ob97} for the
median gas mass ($M_{\text{gas}}\sim10^{9}$~\Mo).

Future wide-field spectroscopy of radio sources, including those of the
low frequency sample, e.g with WEAVE \citep{dta+12,spd+16}, will allow
us to determine many more ionising photon rates for the sources which
have been observed to considerably lower frequencies than the SDSS sample.
These will prove invaluable in determining whether the power-law sources
have simply not been observed to sufficiently low frequencies, as
posited here. With a large number of sources probed to very low
frequencies, we may also be able to constrain which gas distribution
best follows the model and perhaps yield an estimate of the ionising photon
rate in the absence of observed frame UV-optical-IR photometry.
A strong anti-correlation between the abundance of cool neutral gas
and the ionising photon rate has been long established \citep{cww+08,cur24} and,
in the absence of observationally expensive follow-up UV photometry, it will
be interesting to see if the turnover frequency is also anti-correlated
with the detection of \HI\ absorption from ongoing and future surveys
with the SKA and its pathfinders, e.g. the
{\em First Large Absorption Survey in \HI} (FLASH, \citealt{asa+22}). 

\section*{Data availability}

Data available on request and the NED photometry data are included in the online versions.

\section*{Acknowledgements}

I would like to thank the anonymous referee and the scientific editors, Tim Pearson, for
their helpful and constructive comments which helped improve the manuscript.
This research has made use of the NASA/IPAC Extragalactic Database
(NED) which is operated by the Jet Propulsion Laboratory, California
Institute of Technology, under contract with the National Aeronautics
and Space Administration and NASA's Astrophysics Data System
Bibliographic Service. This research has also made use of NASA's
Astrophysics Data System Bibliographic Service and {\sc asurv} Rev 1.2
\citep{lif92a}, which implements the methods presented in
\citet{ifn86}.


\onecolumn
\section*{Appendices}

\subsection*{A: NED photometry}

The NED photometry for each of our samples are included online in the files {\tt SDSS\_NED-fluxes.txt} (the SDSS
sample),
{\tt G23\_NED-fluxes.txt} (the \citeauthor{gsi+23} sample) and {\tt Low-sample\_NED-fluxes.txt}. These are in the format
\begin{verbatim}
No NED1        NED2             z   log10(Freq) log10(Flux)
1  SDSS J205212.82+001137.4 0.6864645 15.29        -4.82
1  SDSS J205212.82+001137.4 0.6864645 15.29        -4.92
1  SDSS J205212.82+001137.4 0.6864645 15.29        -4.83
1  SDSS J205212.82+001137.4 0.6864645 15.13        -4.67
1  SDSS J205212.82+001137.4 0.6864645 15.11        -4.72
\end{verbatim}
where the source number is followed by the NED name, the redshift and $\log_{10}$ of
the observed frequency (Hz) and flux (Jy). Note that {\tt Low-sample\_NED-fluxes.txt} has
an additional field designating the reference for the source: C17 -- \citet{ceg+17},
G22 -- \citet{gds+22}, S22 -- \citet{scr+22}.

\subsection*{B: The \citeauthor{ob97} sample}

In Fig.~\ref{ob97_spectra} we show the SEDs of \citet{ob97} with our fits to the UV (Sect.~\ref{pandf})
and radio (Sect.~\ref{pandf}) photometry.
\begin{figure*}
  \centering\includegraphics[scale=0.42]{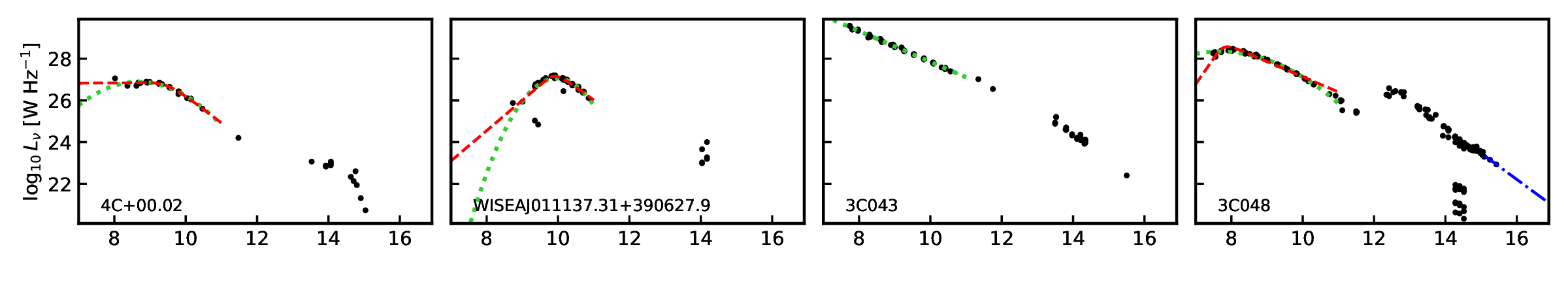}\vspace*{-0.5cm}
  \centering\includegraphics[scale=0.42]{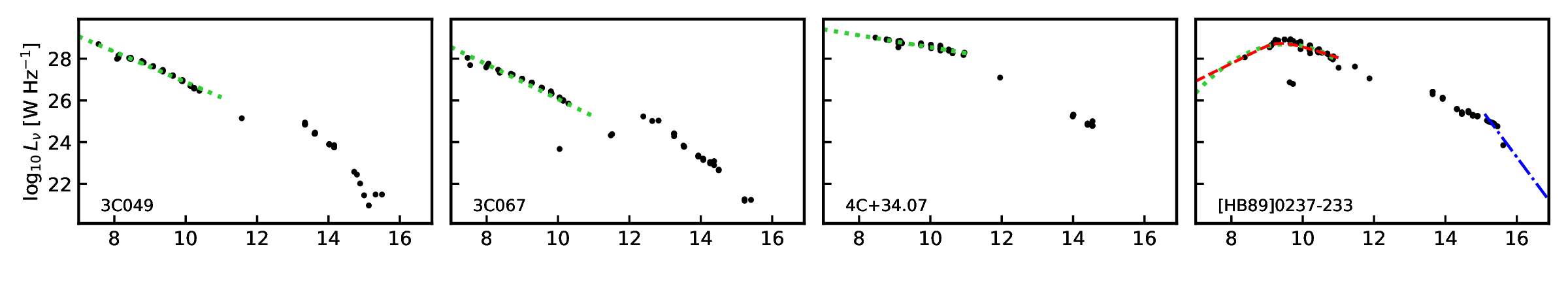}\vspace*{-0.5cm}
  \centering\includegraphics[scale=0.42]{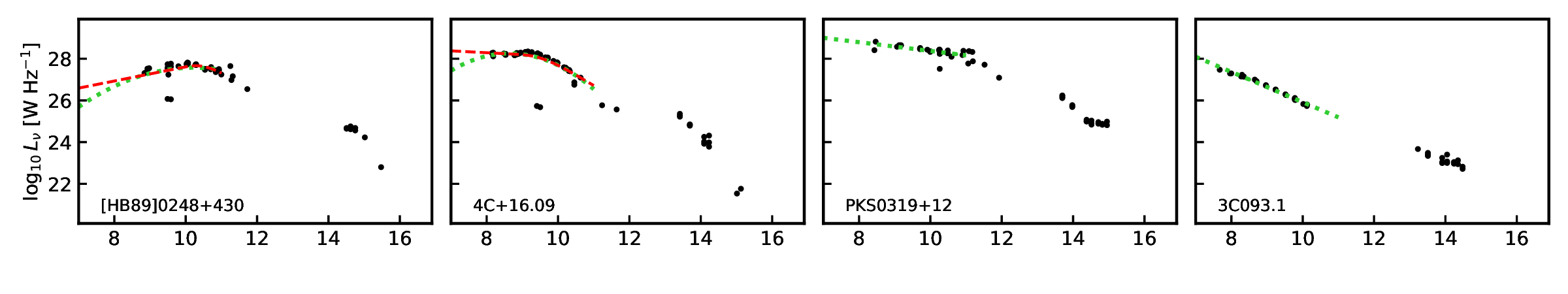}\vspace*{-0.5cm}
  \centering\includegraphics[scale=0.42]{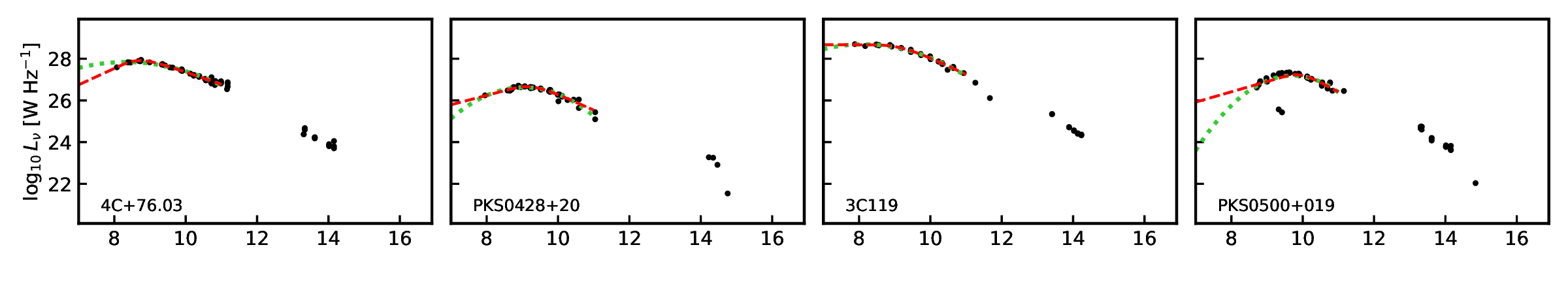}\vspace*{-0.5cm}
  \centering\includegraphics[scale=0.42]{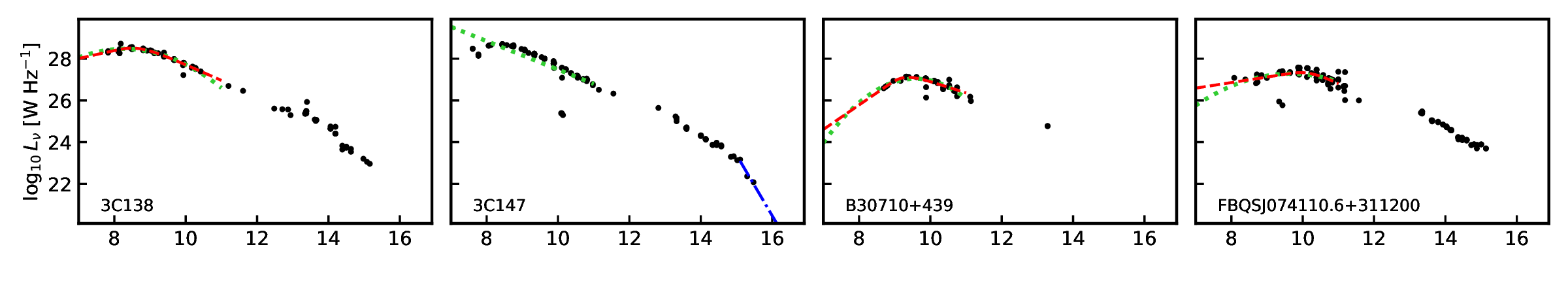}\vspace*{-0.5cm}
  \centering\includegraphics[scale=0.42]{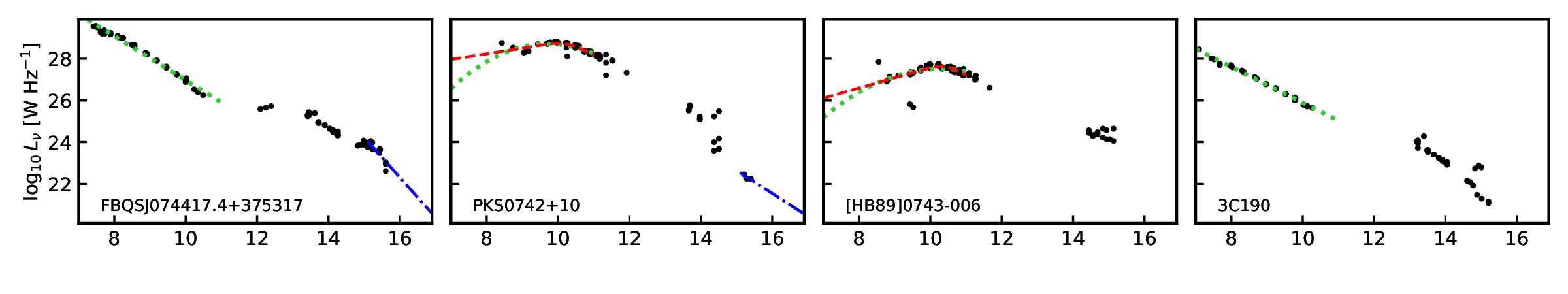}\vspace*{-0.5cm}
  \centering\includegraphics[scale=0.42]{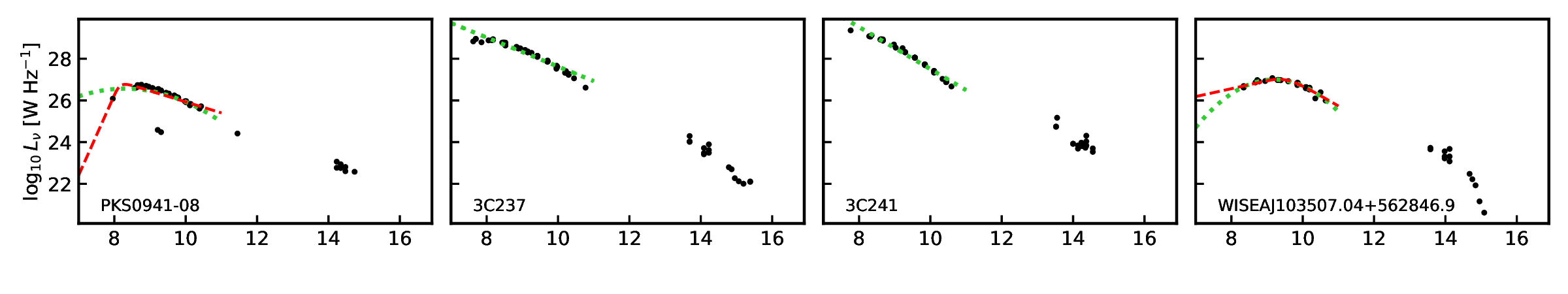}\vspace*{-0.5cm}
  \centering\includegraphics[scale=0.42]{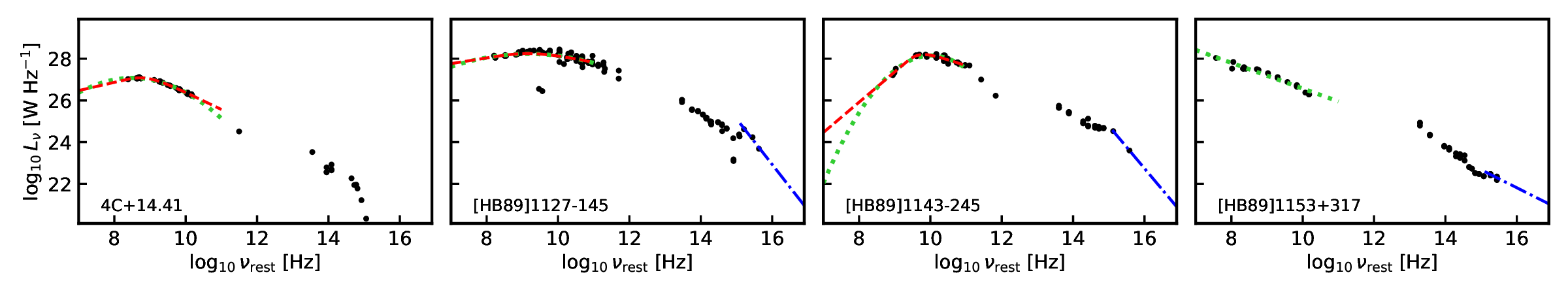}\vspace*{-0.3cm}

  \caption{The SEDs and fits to the \citeauthor{ob97} sample. The curves show the fits to the radio data
  (polynomial --  dotted, GPS-fit -- dashed) and the line the power-law fit to the UV data.}
\label{ob97_spectra}
\end{figure*}

\addtocounter{figure}{-1}
\begin{figure*}
  
  \centering\includegraphics[scale=0.42]{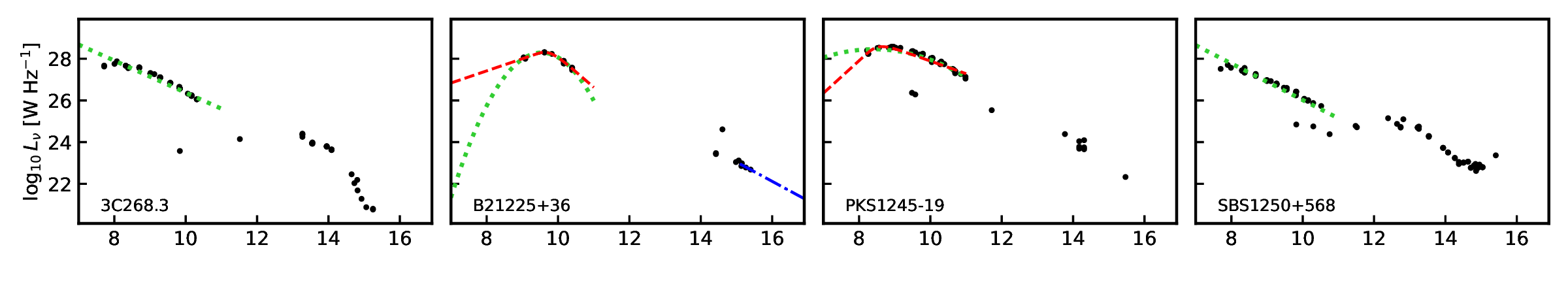}\vspace*{-0.5cm}
  \centering\includegraphics[scale=0.42]{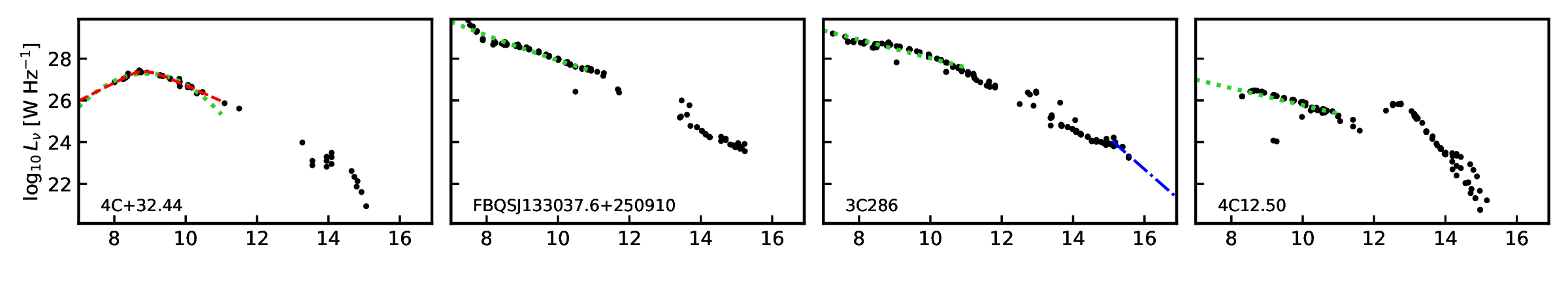}\vspace*{-0.5cm}
  \centering\includegraphics[scale=0.42]{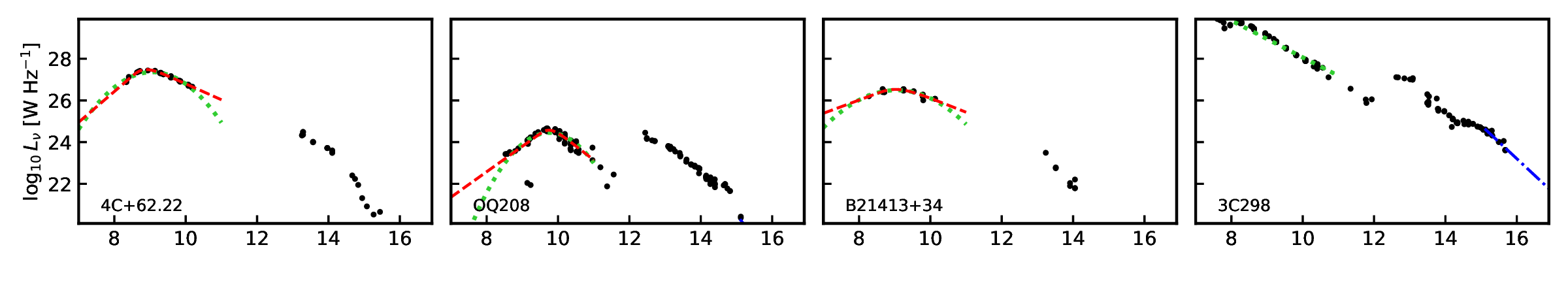}\vspace*{-0.5cm}
  \centering\includegraphics[scale=0.42]{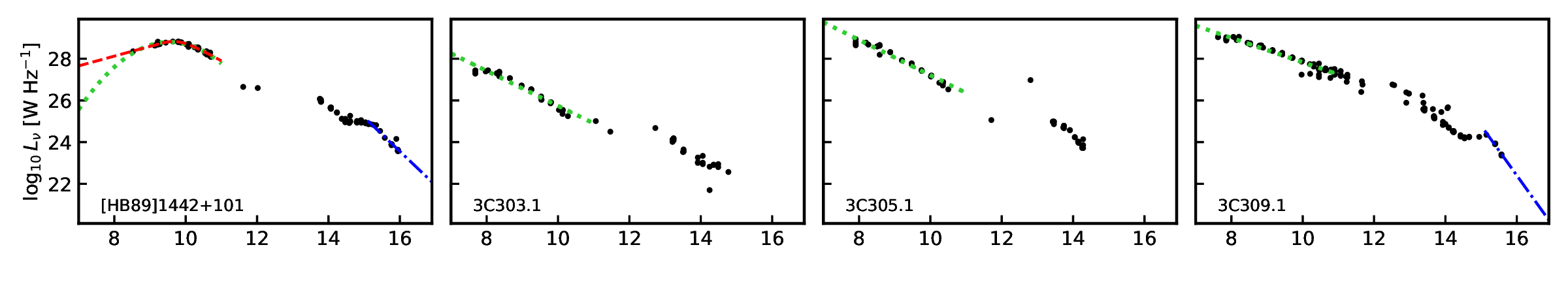}\vspace*{-0.5cm}
  \centering\includegraphics[scale=0.42]{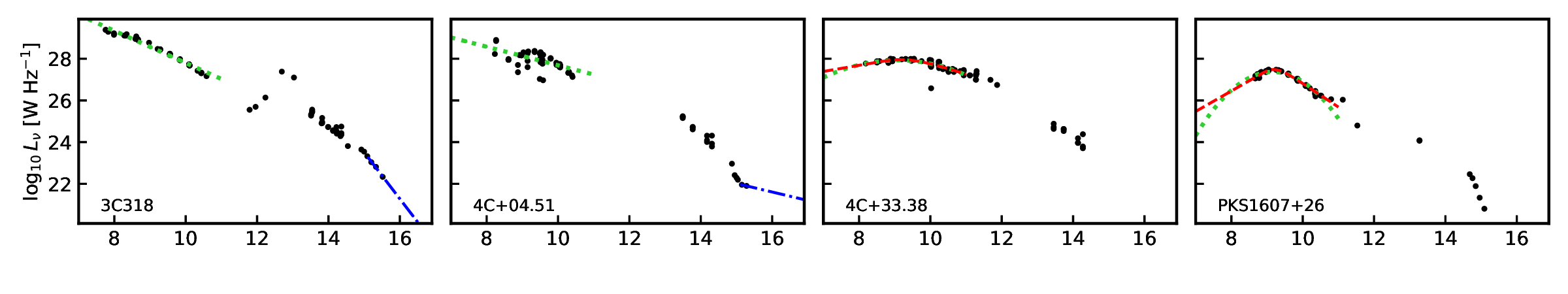}\vspace*{-0.5cm}
  \centering\includegraphics[scale=0.42]{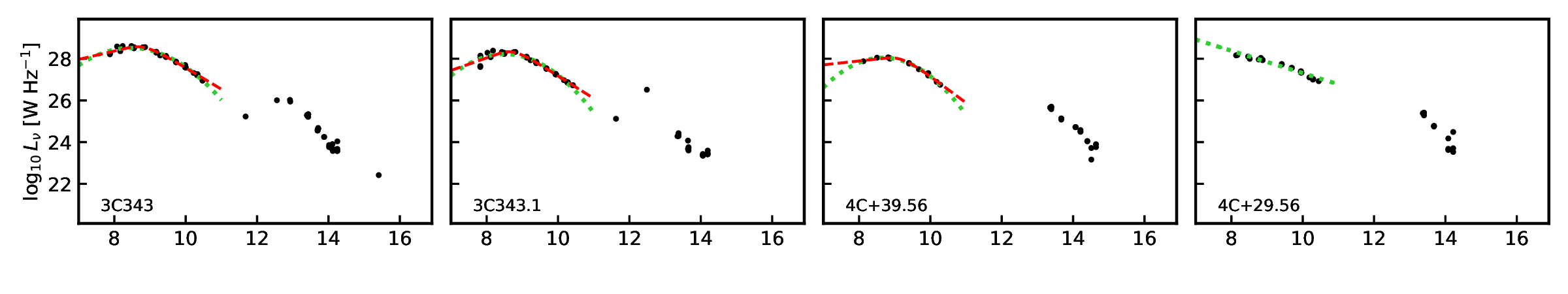}\vspace*{-0.5cm}
  \centering\includegraphics[scale=0.42]{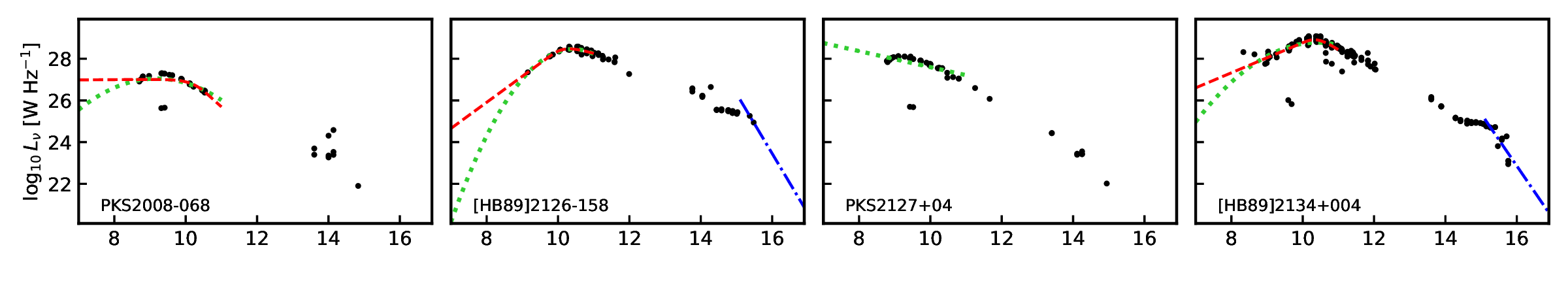}\vspace*{-0.5cm}
  \centering\includegraphics[scale=0.42]{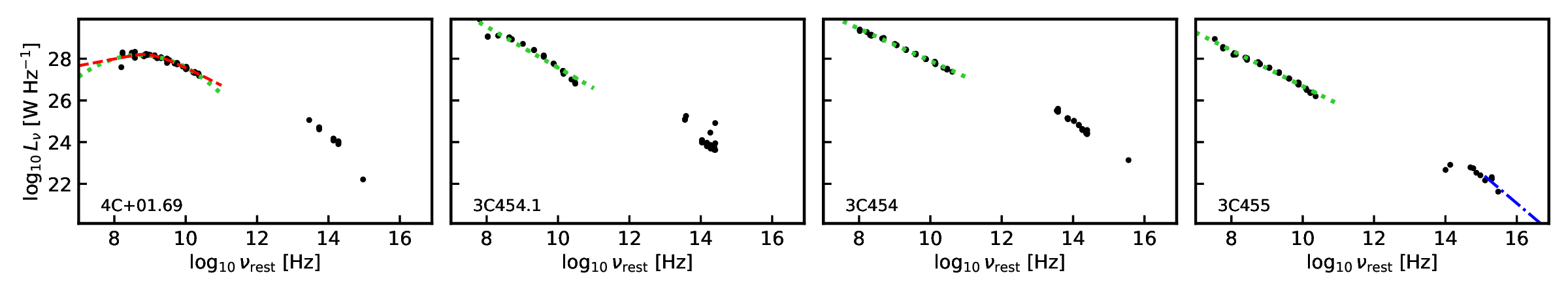}\vspace*{-0.3cm}
  \caption{{\em Continued}}
\end{figure*}

\addtocounter{figure}{-1}
\begin{figure}
  \centering\includegraphics[scale=0.42]{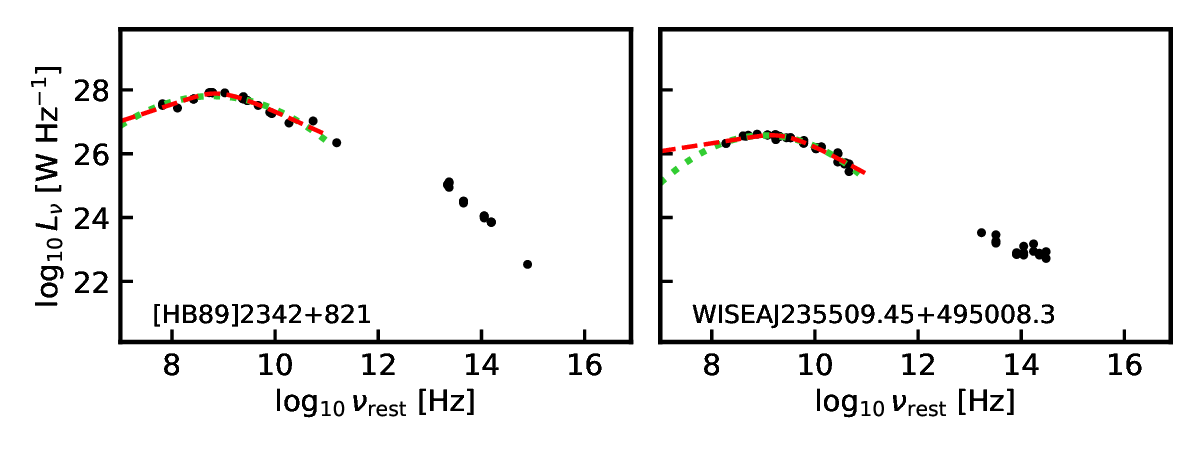}\vspace*{-0.3cm}
\caption{{\em Continued}}
\end{figure}

\begin{figure*}
  \includegraphics[scale=0.52]{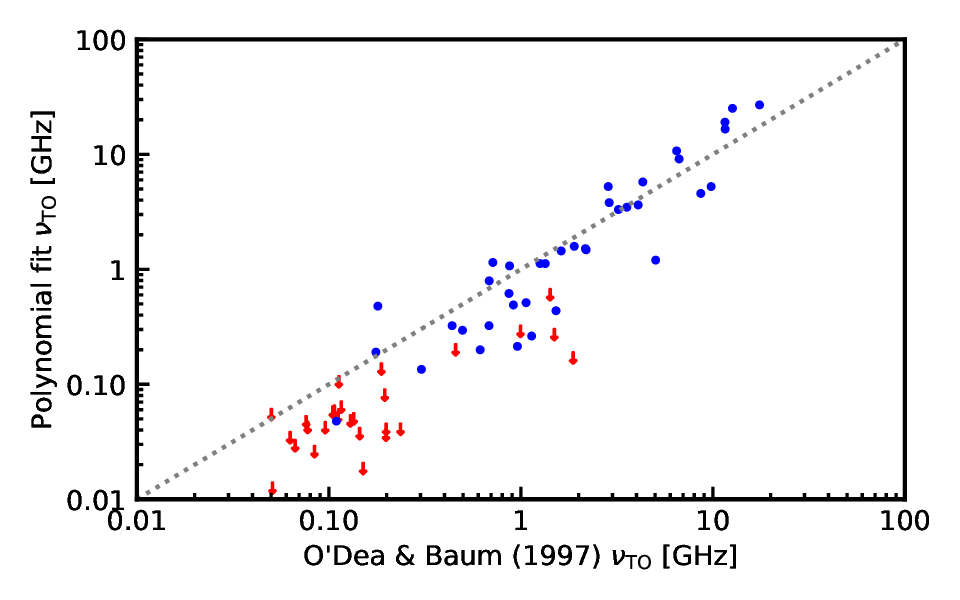} 
  \includegraphics[scale=0.52]{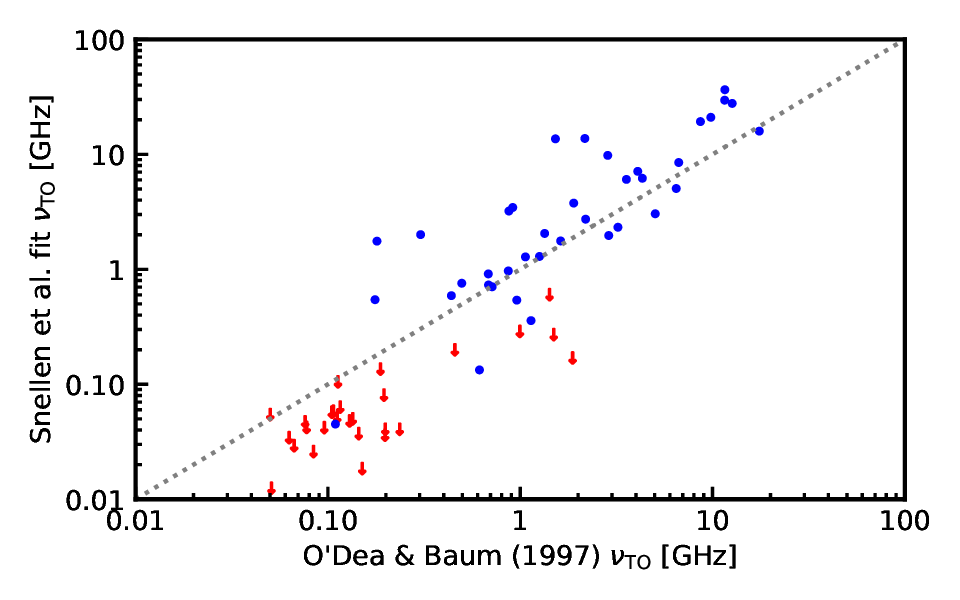}
   \vspace*{-0.3cm}
   \caption{Our derived turnover frequencies compared to those of \citeauthor{ob97}. The arrows show the upper limits and the dotted line where these are equal ($\Delta\log_{10}\nu_{\rm{TO ~[GHz]}}=0$). Left: For the polynomial radio SED fit ($\Delta\log_{10}\nu_{\rm{TO ~[GHz]}}=-0.098$, excluding the limits). Right: For the \citet{ssd+98} radio SED fit ($\Delta\log_{10}\nu_{\rm{TO ~[GHz]}}=0.188$, excluding the limits), see Sect.~\ref{pandf}.}
\label{ob97_TO}
\end{figure*}

In Fig.~\ref{ob97_TO} we show the turnover frequencies obtained from our fits compared to those of \citet{ob97}.

\subsection*{C: The \citeauthor{gsi+23} sample}

\begin{figure*}
\includegraphics[scale=0.42]{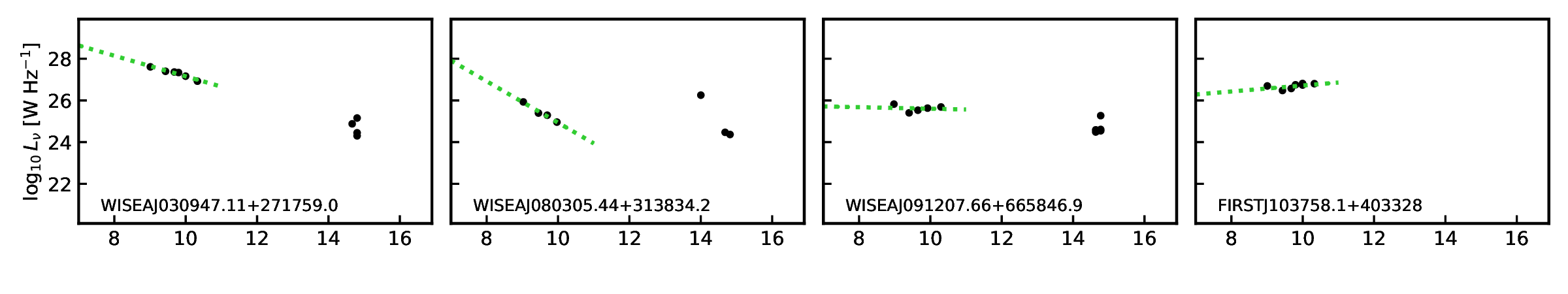}\vspace*{-0.8cm}
\includegraphics[scale=0.42]{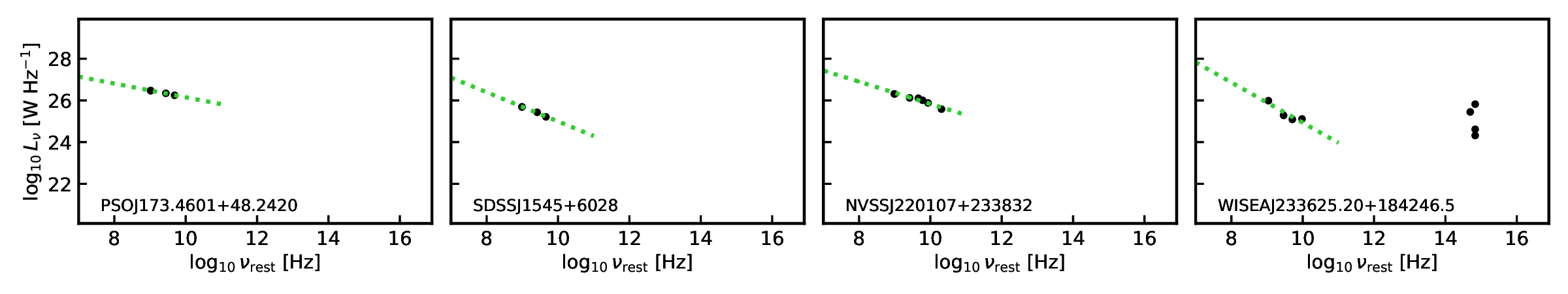}
\caption{The rest-frame photometry of the remaining \citet{gsi+23} sample.}
\label{fits}
\end{figure*}

In addition to Fig.~\ref{J0002}, we show in Fig.~\ref{fits} the
photometry scraped from NED, WISE, 2MASS and GALEX for the sources of
\citet{gsi+23}. Note that for the remaining four sources no photometric
data were available.

\label{lastpage}

\end{document}